\documentclass[conference]{IEEEtran}
\IEEEoverridecommandlockouts
\usepackage{graphicx}
\usepackage{amsmath}
\usepackage{booktabs}
\usepackage{algorithm}
\usepackage{algorithmic}
\usepackage{amsfonts}
\usepackage{multirow}[c]
\usepackage{makecell}
\usepackage{subfigure}
\usepackage{color}
\usepackage{bm}
\usepackage{epstopdf}
\usepackage{url}
\usepackage[cal=cm]{mathalfa}
\usepackage{balance}
\usepackage{threeparttable}
\usepackage{wrapfig}
\usepackage{tabularx}
\usepackage{array}
\usepackage{enumitem}
\usepackage{algorithm}
\usepackage{algorithmic}
\usepackage[dvipsnames]{xcolor}
\usepackage{diagbox}
\usepackage{amssymb}
\usepackage{pifont}
\usepackage[numbers]{natbib}
\setlist[itemize]{leftmargin=*}
\begin{document}

\title{Personalized Multi-Interest Modeling for Cross-Domain Recommendation to Cold-Start Users}

\author{\IEEEauthorblockA{Xiaodong Li\textsuperscript{1,2}, Jiawei Sheng\textsuperscript{1}, Jiangxia Cao\textsuperscript{3}, Xinghua Zhang\textsuperscript{1,2}, Wenyuan Zhang\textsuperscript{1,2}, Yong Sun\textsuperscript{1,2},\\ Shirui Pan\textsuperscript{4}, Zhihong Tian\textsuperscript{5,6,7} and Tingwen Liu\textsuperscript{1,2}$^*$}
\thanks{* Corresponding author.}
\thanks{This work was funded by the Youth Innovation Promotion Association of CAS (No.2021153), the National Natural Science Foundation of China (No.62406319), the Postdoctoral Fellowship Program of CPSF (No.GZC20232968), the Guangdong S\&T Program under Grant 2024B0101010002, the National Natural Science Foundation of China (No.U2436208, No.62372129), and the Project of Guangdong Key Laboratory of Industrial Control System Security (2024B1212020010).}
\IEEEauthorblockA{\textit{\textsuperscript{1}Institute of Information Engineering, Chinese Academy of Sciences, Beijing, China} \\
\textit{\textsuperscript{2}School of Cyber Security, University of Chinese Academy of Sciences, Beijing, China}\\
\textit{\textsuperscript{3}Kuaishou Technology, Beijing, China}\\
\textit{\textsuperscript{4}School of Information and Communication Technology, Griffith University, Queensland, Australia}\\
\textit{\textsuperscript{5}Cyberspace Institute of Advanced Technology, Guangzhou University, Guangdong, China}\\
\textit{\textsuperscript{6}Guangdong Key Laboratory of Industrial Control System Security, Guangdong, China}\\
\textit{\textsuperscript{7}Huangpu Research School of Guangzhou University, Guangdong, China}\\
\{lixiaodong, shengjiawei, zhangxinghua, zhangwenyuan, sunyong, liutingwen\}@iie.ac.cn,\\
jiangxiacao@gmail.com, s.pan@griffith.edu.au, tianzhihong@gzhu.edu.cn}
}



\maketitle

\begin{abstract}
Cross-domain recommendation (CDR) has demonstrated to be an effective solution for alleviating the user cold-start issue.
By leveraging rich user-item interactions available in a richly informative source domain, CDR could improve the recommendation performance for cold-start users in the target domain.
Previous CDR approaches mostly adhere the Embedding and Mapping (EMCDR) paradigm, which learns a user-shared mapping function to transfer users' preference from the source domain to the target domain, neglecting users' \textit{personalized preference}.
Recent CDR approaches further leverage the meta-learning paradigm, considering the CDR task for each user independently and learning user-specific mapping functions for each user. 
However, they mostly learn representations for each user individually, which ignores the \textit{common preference} between different users, neglecting valuable information for CDR. 
In addition, all these approaches usually summarize the user's preference into an overall representation, which can hardly capture the user's \textit{multi-interest preference}. 
To this end, we propose a personalized multi-interest modeling framework for CDR to cold-start users, termed as NF-NPCDR.
Specifically, we propose a personalized preference encoder that enhances the neural process (NP) with the normalizing flow (NF) to convert the Gaussian (unimodal) distribution to a multimodal distribution, providing a novel way to capture the user's personalized multi-interest preference. 
Then, we propose a common preference encoder with a preference pool to capture the common preference between different users. 
Furthermore, we introduce a stochastic adaptive decoder to incorporate both the personalized and common preference for cold-start users, adaptively modulating both preference for better recommendation. 
Experimental evaluations demonstrate that NF-NPCDR outperforms previous SOTA approaches in five benchmark CDR scenarios. 
\end{abstract}



\begin{figure}[t]
\centering
\includegraphics[width=0.95\columnwidth]{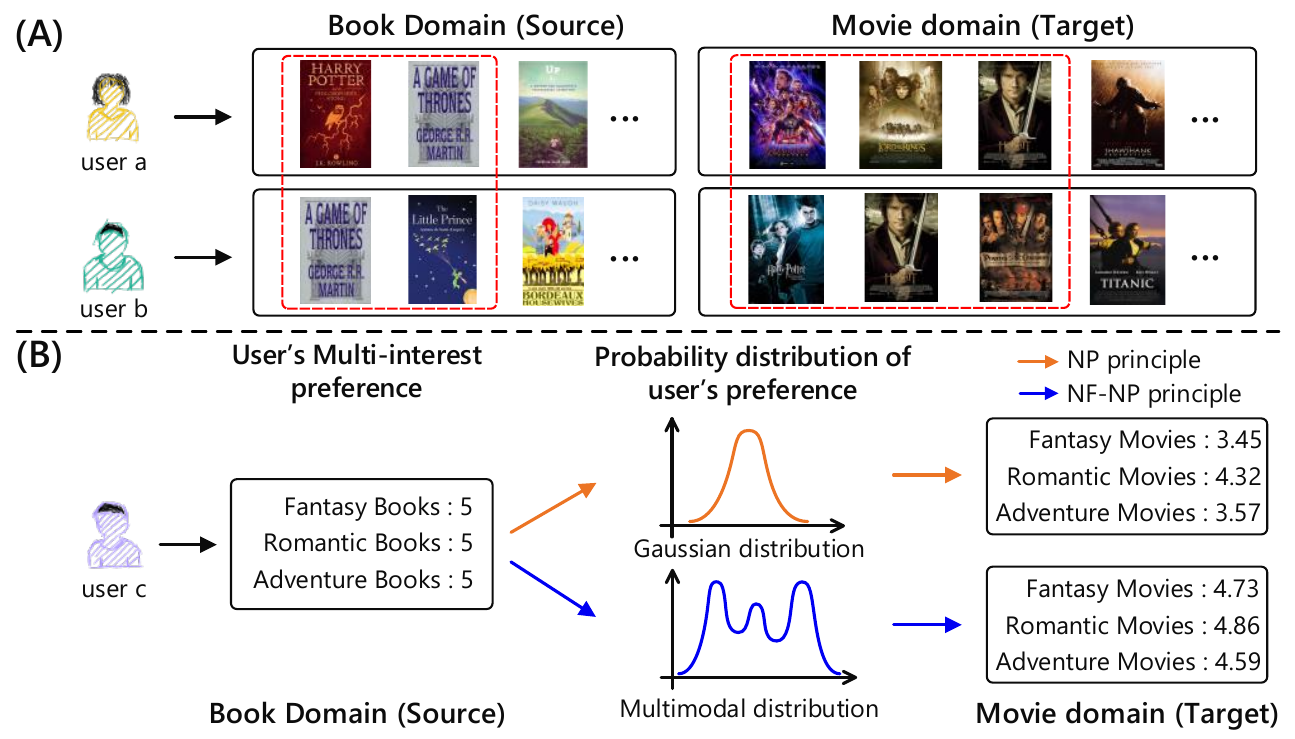} 
\caption{(A) An illustration of the common preference between different users. (B) An illustration of the NP principle and NF-NP principle.}
\vspace{-0.3cm}
\label{motivation}
\end{figure}

\section{Introduction}
Recommendation systems (RS) are extensively applied across various real-world web applications, such as Amazon (e-commerce), Kuaishou (online video) and Twitter (social media).
Collaborative filtering (CF) stands out as an effective approach and has been implemented to RS, including matrix factorization approaches~\cite{BPR,cmf} and modern deep neural network approaches~\cite{ncf,ngcf}.
Nevertheless, CF-based methods typically encounter the persistent user cold-start issue~\cite{tanp}, which makes it difficult to make recommendation for new emerging (cold-start) users with few user-item interactions. Recently, cross-domain recommendation (CDR) has been introduced as an effective solution to the user cold-start issue. Specifically, CDR enhances recommendations for cold-start users in the target domain by exploiting the extensive user-item interactions available in a richly informative source domain.

A mainstream strategy of CDR methods involves training the model with users overlapped between both domains, and subsequently making recommendations for cold-start users in the target domain.
Recent CDR methods~\cite{sscdr,emcdr,tsvcdr,cdrva,tmcdr,dcdcsr,dcdir,hcdir} mostly adhere the Embedding and Mapping (EMCDR~\cite{emcdr}) paradigm, which first generates user/item embeddings from both domains respectively, and then learns a mapping function to transfer users' preference from the source domain to the target domain.
However, these EMCDR-based methods usually learn a user-shared mapping function for all users, which neglects users' \textit{\textbf{personalized preference}}, thus limiting the transfer effectiveness for CDR. 

To capture the personalized preference of users, some meta-learning-based CDR methods~\cite{tmcdr,ptupcdr,cdml} have been proposed, which consider the CDR task for each user independently, and thus learn user-specific mapping functions for each user by meta-learning paradigm.
Typically, PTUPCDR~\cite{ptupcdr} proposes a meta-network to generate the personalized mapping functions based on users representations to transfer the personalized preference for each user. 
However, these meta-learning-based methods learn representations for each user individually, which ignores the common preference between different users.
In contrast, we recognize that the users can have \textbf{\textit{common preference}}, which provides crucial information for CDR. 
For example in Fig.~\ref{motivation}(A), both user \textit{a} and \textit{b} read fantasy books in the Book (source) domain, reflecting a common preference namely \textit{fantasy-interest}.
Besides, we can also observe that both users watch fantasy movies in the Movie (target) domain, exhibiting the same preference in the target domain. 
Therefore, learning the common preference between different users could acquire abundant information from their related users, thus benefit user's preference modeling.


Although the above methods have achieved excellent results, these EMCDR-based and meta-learning-based methods typically summarize user's preference as an overall representation, which can hardly reflect the user's \textbf{\textit{multi-interest preference}}~\cite{cmiffr, epimi}.
As shown in Fig.~\ref{motivation}(B), user \textit{c} has multiple interests in the Book (source) domain, namely \textit{fantasy-interest}, \textit{romantic-interest} and \textit{adventure-interest}.
Directly learning the user preference with a simple and deterministic representation may bias part of the user's interests and lead to unitary recommendation in target domain.

To this end, we seek the unification of neural process (NP)~\cite{np} and normalizing flow (NF)~\cite{nf,ndea} as a promising solution.
In general, NP provides a neural-based approximation of stochastic processes, which can model uncertain distributions over prediction functions. As shown in Fig.~\ref{motivation}(B), NP maps user's preference to a Gaussian distribution, and extends the meta-learning paradigm to transfer the user's personalized preference from the source to target domain.
Besides, we further enhance neural process with the normalizing flow, which can convert the Gaussian (unimodal) distribution to a multimodal distribution, thus helping to capture the user's personalized multi-interest preference.
In this way, the unification of NP and NF allows for learning more powerful CDR mapping functions.


Following the above idea, we propose a novel personalized multi-interest modeling framework for CDR to cold-start users, termed as NF-NPCDR.
Specifically, we propose a personalized preference encoder that enhances the neural process with the normalizing flow to capture the user's personalized multi-interest preference. 
Then, we propose a common preference encoder with a preference pool to capture the common preference between different users. 
Furthermore, we devise a stochastic adaptive decoder to incorporate both the personalized and common preference for cold-start users, adaptively modulating both preference for better recommendation.
To summarize the main contributions of this paper:
\begin{itemize}
    \item We introduce a multi-interest modeling framework to address the cold-start issue in CDR with the unification of neural process and normalizing flow.
    \item We propose a personalized preference encoder to capture the user's personalized multi-interest preference. Then we propose a common preference encoder to capture the common preference between different users. A stochastic adaptive decoder is also designed to further incorporate both the personalized and common preference for cold-start users.
    \item Extensive experiments demonstrate the effectiveness and superiority of our NF-NPCDR in five CDR scenarios. Further experiments demonstrate the significant ability of NF-NPCDR to capture user's multi-interest preference and model the common preference between different users.
\end{itemize}

The rest of this paper is organized as follows. A summary of related works is shown in Section~\ref{related_works}. We briefly review the neural process and normalizing flow techniques, as well as the problem definition in Section~\ref{preliminary}. Section~\ref{approach} shows the personalized multi-interest modeling framework of our model. We present the experimental results and analyses of our NF-NPCDR in Section~\ref{experiments}.
We finally conclude our approach and give the future work in Section~\ref{conclusion}.

\section{Related Work}\label{related_works}


\textbf{Cross-domain recommendation}~\cite{dlcdr,abrccdr,dmlee,cdrpsa} aims to improve the performance of recommendation to users in the target domain by utilizing rich auxiliary information from the source domain. 
Recent studies of CDR can generally be categorized into two distinct types based on the aiming user groups: one focuses on addressing the data sparsity issue~\cite{conet,cmf,cdcfsr,disencdr} for overlapping users, and the other concentrates on solving the cold-start issue~\cite{emcdr,sscdr,dcdir,cdml,cpkt} for cold-start (i.e., non-overlapping) users.
In this article, we focus on the cold-start issue in the CDR task.

To alleviate the cold-start issue, several CDR methods~\cite{emcdr,dcdcsr,sscdr,dcdir,hcdir} following Embedding and Mapping paradigm have been proposed. Specifically, EMCDR~\cite{emcdr} learns a mapping function between the source and target domains to transfer users' preference. SSCDR~\cite{sscdr} proposes a semi-supervised mapping function to learn the cross-domain relationship. DCDCSR~\cite{dcdcsr} employs the MF models and DNN to map the user and item latent factors across domains or systems. DCDIR~\cite{dcdir} and HCDIR~\cite{hcdir} construct heterogeneous information networks and leverage the user-item interactions to learn the mapping function. Recent several meta-learning-based methods~\cite{tmcdr,ptupcdr} also follow the Embedding and Mapping paradigm. For example, TMCDR~\cite{tmcdr} and PTUPCDR~\cite{ptupcdr} propose a meta network to transfer the personalized preference of users. More recently, CDRIB~\cite{cdrib} attempts to learn disentanglement representations via information bottleneck. 
REMIT~\cite{remit} leverages user's multiple interests by multiple deterministic interest embeddings, but neglects the uncertainty modelling of user interests.
Our NF-NPCDR belongs to the meta-learning-based method and has significant design differences with the above methods that we capture the user's personalized multi-interest preference and the common preference between different users.

\textbf{Neural Process}~\cite{np} combines the strengths of Gaussian process and neural networks to define distributions over functions and can estimate the uncertainty in their predictions. 
An increasing amount of researches~\cite{cnp,np,anp} have been proposed to enhance the expressiveness of the NP models.
CNP~\cite{cnp} make predictions with only few training data points, and can be extended to large datasets.
NP~\cite{np} is proposed to learn the latent random variable and can estimate the uncertainty of predictions. ANP~\cite{anp} incorporates attention into NP to address the underfitting issue. 
The characteristics of NP have been applied on multi-task learning~\cite{mtlnp} and image recognition~\cite{NP-SemiSeg}.
Recently, NP has been applied to the field of recommendation systems~\cite{idnp,tanp,tfanp,lieinp,cdrnp}. IDNP~\cite{idnp} leverages dilated convolutions and neural process to model user's short-term and long-term interests. CDRNP~\cite{cdrnp} leverages attention mechanism to bridge the source and target domains, and the neural process to capture the preference correlations among users. TANP~\cite{tanp} follows the meta-learning paradigm to make predictions for new tasks with neural process.
INP~\cite{lieinp} focuses on capturing user's diverse intentions from intrinsic-level with neural stochastic process. Nevertheless, NF-NPCDR directly transfers users' personalized preference from source domain to target domain via neural process, having a fundamentally different framework from the above methods.


\textbf{Normalizing flows}~\cite{nf,ndea} enable to convert a Gaussian (unimodal) distribution to a multimodal distribution by applying a sequence of invertible and differentiable mappings. There have been numerous works~\cite{nfs} applying normalizing flows in the field of statistics and machine learning. Planar and Radial flows~\cite{planarflow} are used to approximate posterior distributions, which are favored for their simplicity and computational ease. These two flows are only suitable for low-dimensional situations. In order to deal with high-dimensional and highly structured data, RealNVP~\cite{realnvpflow} is a suitable choice, which uses real-valued non-volume preserving transformations. MAF~\cite{mafflow} generalizes RealNVP by building a sequence of autoregressive models, resulting in a type of normalizing flow well-suited for density estimation.
Normalizing flows have been applied to many tasks, such as reinforcement learning~\cite{lenf,rvfnf}, graph generation~\cite{graphnvp} and graph completion~\cite{npfkgc}. In this paper, we further enhance neural process with the normalizing flow to model the user's personalized multi-interest preference for CDR to cold-start users.



\section{Preliminary and problem definition}\label{preliminary}
This section first introduces the background knowledge of the neural process and normalizing flow, then formally defines our CDR problem settings.

\subsection{Preliminary}

\subsubsection{Neural Process}

Neural Process (NP)~\cite{np,cnp} represents a class of neural latent variable models that approximate stochastic processes $f: X \to Y$ using neural networks with robust parameterization capabilities.
Consider a set of dataset $\mathcal{T} = \{x_i, y_i\}_{i=1}^{|\mathcal{T}|}$, the corresponding probability distribution over every $(x_i, y_i)$ can be defined as follows:
\begin{equation}
\small
\mathbf{\rho}_{x_{1:|\mathcal{T}|}}(y_{1:|\mathcal{T}|}) = \int p(y_{1:|\mathcal{T}|}|x_{1:|\mathcal{T}|},f)p(f)df,
\label{}
\end{equation}


NP leverages neural networks (NNs) with a latent random variable $\bm{z}$ to parameterize the stochastic processes $f$. Specifically, NP splits the dataset $\mathcal{T}$ into a support set $\mathcal{C} = \{x_i, y_i\}_{i=1}^{|\mathcal{C}|}$ and a query set $\mathcal{Q} = \{x_i, y_i\}_{i=1}^{|\mathcal{Q}|}$. Given the support set $\mathcal{C}$, NP aims to make predictions on the query set $\mathcal{Q}$ with $p(\bm{z}|\mathcal{C})$ derived from $\mathcal{C}$. Consequently, we rewrite the Bayesian inference as follows:
\begin{equation}
\small
\begin{split}
p(y_{1:|\mathcal{Q}|}|x_{1:|\mathcal{Q}|},\mathcal{C}) = \int p(y_{1:|\mathcal{Q}|}|x_{1:|\mathcal{Q}|},\bm{z})p(\bm{z}|\mathcal{C})d\bm{z},
\end{split}
\label{}
\end{equation}
Here we assume $p(\bm{z}|\mathcal{C})\sim \mathcal{N}(\bm{0},\bm{I})$ for simplicity in paradigmatic variational frameworks~\cite{aevb}. However, the true posterior distribution is intractable when optimising the model parameters. Thus, NP adopts amortized variational inference~\cite{aevb,nvi} to derive the corresponding evidence lower-bound (ELBO) function as follows:
\begin{equation}
\small
\begin{split}
&\mathrm{log}p(y_{1:|\mathcal{Q}|}|x_{1:|\mathcal{Q}|},\mathcal{C}) = \mathrm{log}\int p(\bm{z},y_{1:|\mathcal{Q}|}|x_{1:|\mathcal{Q}|},\mathcal{C})d\bm{z}\\
&\geqslant  E_{q(\bm{z}|\mathcal{C},\mathcal{Q})}\left[\log\frac{p(\bm{z},y_{1:|\mathcal{Q}|}|x_{1:|\mathcal{Q}|},\mathcal{C})}{q(\bm{z}|\mathcal{C},\mathcal{Q})}\right]\\
&= E_{q(\bm{z}|\mathcal{C},\mathcal{Q})}\left[\sum_{i=1}^{|\mathcal{Q}|}{\log p(y_i|x_i,\bm{z})}+\log\frac{p(\bm{z}|\mathcal{C})}{q(\bm{z}|\mathcal{C},\mathcal{Q})}\right]\\
&=\underbrace{E_{q(\bm{z}|\mathcal{C},\mathcal{Q})}\!\log p(y_{1:|\mathcal{Q}|}|x_{1:|\mathcal{Q}|},\bm{z})}_{\text{desired log-likelihood}} - \underbrace{\mathrm{KL}(q(\bm{z}|\mathcal{C},\mathcal{Q})||p(\bm{z}|\mathcal{C}))}_{\text{KL divergence}},
\end{split}
\label{np_loss}
\end{equation}
The derivation results of Eq.~(\ref{np_loss}) can be clearly understood in terms of two components: the first focuses on achieving the desired CDR task, while the second ensures that the support set $\mathcal{C}$ and the query set $\mathcal{Q}$ following the same stochastic process.

\subsubsection{Normalizing Flows}

Normalizing Flows (NF)~\cite{nf,vinf,npfkgc,ndea} are a series of invertible and differentiable mappings used to convert probability distributions. Specifically, they convert the Gaussian (unimodal) distribution to a multimodal distribution. In the framework of NF, $\bm{z}_0$ is drawn from the original distribution (e.g., $q_0(\bm{z}_0|\mathcal{C},\mathcal{Q})$). The transformation of the original distribution is governed by an invertible and differentiable mappings, denoted as $g_k: \mathbb{R}^{d} \to \mathbb{R}^{d}$. 
By applying a sequence of $g_k$ with $K$ steps, the final random variable $\bm{z}_K$ is obtained as follows:
\begin{equation}
\small
\begin{split}
\bm{z}_K &= g_K \circ ... \circ g_2 \circ g_1(\bm{z}_0), \\
q_K(\bm{z}_K|\mathcal{C},\mathcal{Q}) &= q_0(\bm{z}_0|\mathcal{C},\mathcal{Q})\prod_{k=1}^{K}\bigg|det\frac{\partial g_k}{\partial \bm{z}_{k-1}}\bigg|^{-1},
\end{split}
\label{nf_loss}
\end{equation}
where $q_K(\bm{z}_K|\mathcal{C},\mathcal{Q})$ is the distribution of the final random variable $\bm{z}_K$, and $|det\frac{\partial g_k}{\partial \bm{z}_{k-1}}|^{-1}$ is the determinant of the Jacobian of $g_k$ at random variable $\bm{z}_{k-1}$.

\begin{figure}[t]
\centering
\includegraphics[width=0.95\columnwidth]{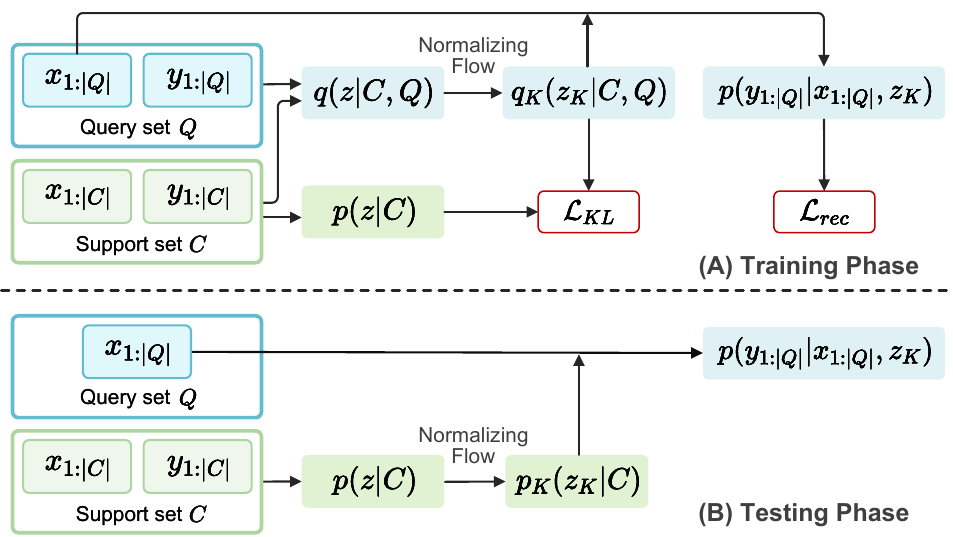} 
\caption{An illustration of our NF-NPCDR during the training phase and testing phase implementing the NF-NP principle.}
\vspace{-0.3cm}
\label{NF_NP_principle}
\end{figure}

\subsubsection{Unification of Neural Process and Normalizing Flow}

In this paper, we consider to further enhance the neural process with normalizing flow. By integrating Eq.~(\ref{nf_loss}) into Eq.~(\ref{np_loss}), we refine the ELBO function as follows:
\begin{equation}
\small
\begin{split}
&E_{q(\bm{z}|\mathcal{C},\mathcal{Q})}\!\log p(y_{1:|\mathcal{Q}|}|x_{1:|\mathcal{Q}|},\bm{z}) - \mathrm{KL}(q(\bm{z}|\mathcal{C},\mathcal{Q})||p(\bm{z}|\mathcal{C})) \\
&\simeq E_{q_0(\bm{z}_0|\mathcal{C},\mathcal{Q})}\!\left[\log p(y_{1:|\mathcal{Q}|}|x_{1:|\mathcal{Q}|},\bm{z}_K)-\mathrm{log}\frac{q_K(\bm{z}_K|\mathcal{C},\mathcal{Q})}{p(\bm{z}_K|\mathcal{C})}\right] \\
&=\underbrace{E_{q_0(\bm{z}_0|\mathcal{C},\mathcal{Q})}\!\log p(y_{1:|\mathcal{Q}|}|x_{1:|\mathcal{Q}|},\bm{z}_K)}_{\text{desired log-likelihood}} \\
& -\underbrace{E_{q_0(\bm{z}_0|\mathcal{C},\mathcal{Q})}\bigg[\log q_0(\bm{z}_0|\mathcal{C},\mathcal{Q}) - \sum_{k=1}^{K}\bigg|\mathrm{det}\frac{\partial g_k}{\partial \bm{z}_{k-1}}\bigg| - \log p(\bm{z}_K|\mathcal{C})\bigg]}_{\text{KL divergence}},
\end{split}
\label{nf_np_loss}
\end{equation}

In terms of implementation, we demonstrate the NF-NP paradigm in a practical standpoint. Fig.~\ref{NF_NP_principle} displays NF-NPCDR employing the NF-NP principle during both the training phase and testing phase. Specifically, during the training phase, the normalizing flow is adopted to convert the Gaussian (unimodal) distribution $q(\bm{z}|\mathcal{C},\mathcal{Q})$ to a multimodal distribution $q_K(\bm{z}_K|\mathcal{C},\mathcal{Q})$, and then we use $\mathcal{L}_{KL}$ in Eq.~(\ref{nf_np_loss}) to assimilate $q_K(\bm{z}_K|\mathcal{C},\mathcal{Q})$ and $p(\bm{z}|\mathcal{C})$. During the testing phase, the ground truth $y_{1:|\mathcal{Q}|}$ in query set $\mathcal{Q}$ is not available. We first use normalizing flow to convert the $p(\bm{z}|\mathcal{C})$ to $p_K(\bm{z}_K|\mathcal{C})$, then $\bm{z}_K$ sampled from the $p_K(\bm{z}_K|\mathcal{C})$ is substituted for $q_K(\bm{z}_K|\mathcal{C},\mathcal{Q})$, and thus together with $x_{1:|\mathcal{Q}|}$ in the query set, our NF-NPCDR make recommendations by $p(y_{1:|\mathcal{Q}|}|x_{1:|\mathcal{Q}|},\bm{z}_K)$.

\begin{table}[t]
\centering
\footnotesize
\caption{Notations.}
\setlength{\tabcolsep}{2pt}
\label{notations}
\begin{tabular*}{0.45 \textwidth}{@{\extracolsep{\fill}}@{}cl@{}}
\toprule
\textbf{Notation} & \textbf{Description}\\
\midrule
$\mathcal{D}^s$ and $\mathcal{D}^t$  & Interaction data \\
$\mathcal{U}^s$ and $\mathcal{U}^t$ &  User set \\
$\mathcal{V}^s$ and $\mathcal{V}^t$  &  Item set \\
$\mathcal{Y}^s$ and $\mathcal{Y}^t$ &  Rating matrix \\
$\mathcal{U}^{s\setminus o}$  &  Cold-start (i.e., non-overlapping) user set \\
$\mathcal{U}^o$  &  Overlapping user set across domains \\
$\mathcal{T}_i$, $\mathcal{C}_i$ and $\mathcal{Q}_i$  &  Task, Support set and Query set for user $u_i$\\
$\Omega^{tr}$ and $\Omega^{te}$ &  Training task set and Testing task set\\
$|\mathcal{C}_i|$ and $|\mathcal{Q}_i|$ &  Number of interactions in $\mathcal{C}_i$ and $\mathcal{Q}_i$ \\
$y_{i,j}^s$ and $y_{i,j}^t$ &  Actual rating score of user $u_i$ to item $v_j^s$ and $v_j^t$\\ 
$\hat{y}_{i,j}^t$  &  Prediction rating score of user $u_i$ to item $v_j^t$ \\
\midrule
$d_1$, $d_2$, $d_3$ & Embedding dimension \\
$\bm{\mu}_i$  & Mean of the Gaussian distribution \\
$\bm{\sigma}_i$  & Variance of the Gaussian distribution \\
$\bm{\epsilon}$  &  Gaussian noise \\
$g_K$  &  Invertible transformation function \\
$K$  &   Normalizing flow step \\
$\mathcal{P}$ &  Preference pool \\ 
$N$  & Number of the soft cluster centroids \\
$\mathcal{M}$ &  Soft assignments matrix \\ 
$\mathcal{D}$ &  Auxiliary distribution \\
$\bm{\eta}^{l}_{i}$ and $\bm{\delta}^{l}_{i}$ &  Modulation parameters of stochastic adaptive \\ & decoder for the $l$-th layer \\
\midrule
$\lambda$  &  Hyper-parameter to balance $\mathcal{L}_c$ \\
$\mathcal{L}_c$ &  KL loss between soft assignments $\mathcal{M}$ and \\ & auxiliary distribution $\mathcal{D}$  \\
$\mathcal{L}_{\mathrm{KL,i}}$  & KL loss rewritten with the normalizing flow \\
$\mathcal{L}_{rec,i}$  &  Desired log-likelihood for rating prediction \\
$\mathcal{L}$  &  Overall loss function \\

\bottomrule
\end{tabular*}
\end{table}

\subsection{Problem Definition}

This study considers a general CDR scenario with the source domain and the target domain. Let $\mathcal{D}^s = (\mathcal{U}^s, \mathcal{V}^s, \mathcal{Y}^s)$ and  $\mathcal{D}^t = (\mathcal{U}^t, \mathcal{V}^t, \mathcal{Y}^t)$ denote the interaction data from source domain and target domain, where $\mathcal{U}$, $\mathcal{V}$ and $\mathcal{Y}\in \{0,1,2,3,4,5\}^{|\mathcal{U}|\times |\mathcal{V}|}$ represent the user set, item set and rating matrix. Here $y_{i,j}\in \mathcal{Y}$ is the rating score, which indicates the user $u_i\in \mathcal{U}$ preference for selecting a particular item $v_j\in \mathcal{V}$. Specifically, the user sets $\mathcal{U}^s$ and $\mathcal{U}^t$ include an overlapping subset of users, referred to as $\mathcal{U}^o = \mathcal{U}^s \cap ~\mathcal{U}^t$, then we leverage $\mathcal{U}^{s\setminus o} = \mathcal{U}^{s}\setminus \mathcal{U}^o$ to represent the cold-start (i.e., non-overlapping) users that exist only in the source domain.

Consequently, given the user-item interactions from both the source and target domains, CDR aims to predict the rating scores for cold-start users in the target domain by exploiting the extensive user-item interactions in the source domain.
Formally, we define our CDR problem with task $\mathcal{T}_i$, which aims to make personalized recommendation for user $u_i \in \mathcal{U}$. The concept of a task $\mathcal{T}_i$ includes user's interactions from source domain as the support set $\mathcal{C}_i$ and interactions from the target domain as the query set $\mathcal{Q}_i$, i.e., $\mathcal{T}_i = \mathcal{C}_i\cup \mathcal{Q}_i$.
Here we denote $\mathcal{C}_i = \{u_i,v_j^s, y_{i,j}^s\}_{j=1}^{|\mathcal{C}_i|}$ and $\mathcal{Q}_i = \{u_i,v_j^t, y_{i,j}^t\}_{j=1}^{|\mathcal{Q}_i|}$.

For the above purpose, we train our model on the training tasks $\Omega^{tr}$ with overlapping users $u_i^o\in \mathcal{U}^o$, i.e., we use $\mathcal{C}_i$ from source domain and $\mathcal{Q}_i$ from target domain to learn the cross domain knowledge. In the testing phase, our model makes recommendation to cold-start users $u_i^{s\setminus o}\in \mathcal{U}^{s\setminus o}$ for new testing tasks $\Omega^{te}$ with only $\mathcal{C}_i$ available, i.e., we use $\mathcal{C}_i$ from source domain and the cross domain knowledge learned from $\Omega^{tr}$ to predict the rating score $y^t_{i,j}$ for each candidate item $v^t_j \in \mathcal{V}^t$ in $\mathcal{Q}_i$ from the target domain. 
The notations used in this paper are summarized in Table \ref{notations}.

\begin{figure*}[t!]
\begin{center}
\includegraphics[width=17.5cm]{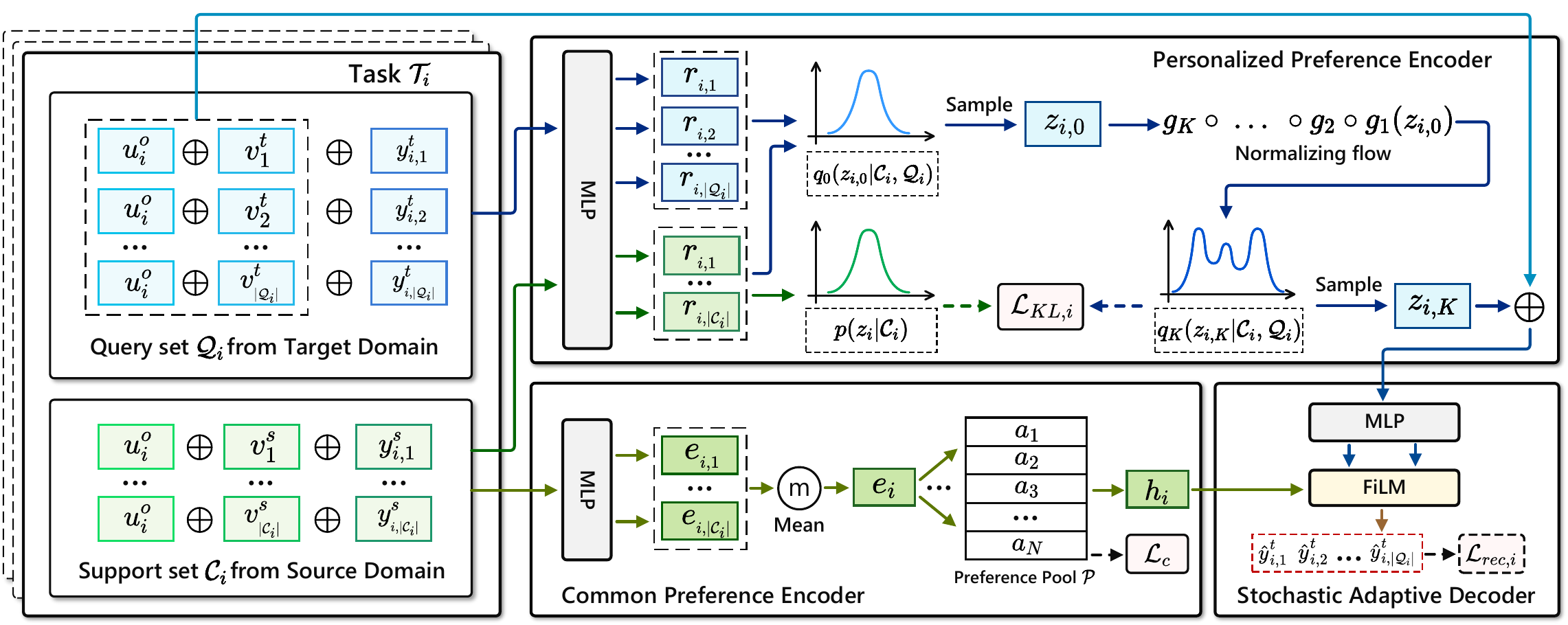}
\caption{Overview of NF-NPCDR, including personalized preference encoder, common preference encoder and stochastic adaptive decoder. Both of $\mathcal{C}_i$ and $\mathcal{T}_i$ are encoded by the neural process encoder to generate the $p(\bm{z}_i|\mathcal{C}_i)$ and $q_0(\bm{z}_{i,0}|\mathcal{C}_i,\mathcal{Q}_i)$. The $q_0(\bm{z}_{i,0}|\mathcal{C}_i,\mathcal{Q}_i)$ is further converted by the normalizing flow encoder to generate the $q_K(\bm{z}_{i,K}|\mathcal{C}_i,\mathcal{Q}_i)$. $\mathcal{C}_i$ is encoded by the common preference encoder to generate the final preference representation $\bm{h}_i$. Finally, $\bm{h}_i$ is used as input to the FiLM and $\bm{z}_{i,K}$ is concatenated with $(\bm{u}_i^o,\bm{v}_j^t)$ in $\mathcal{Q}_i$ to predict $\hat{y}_{i,j}^t$ via stochastic adaptive decoder.}
\vspace{-0.3cm}
\label{main_model}
\end{center}
\end{figure*}

\section{Approach}\label{approach}

In this section, we present our proposed approach NF-NPCDR, which consists of four components: 1) An embedding layer to generate initialized user/item representations as the inputs of NF-NPCDR; 2) A personalized preference encoder to capture the user's personalized multi-interest preference; 3) A common preference encoder to capture the common preference between different users; 4) A stochastic adaptive decoder to incorporate both the personalized preference and common preference for cold-start users, ultimately generating predicted rating scores for candidate items in the target domain. The framework of NF-NPCDR is illustrated in Fig.~\ref{main_model}.

\subsection{Embedding Layer}

The embedding layer embeds the representations of users and items into a low dimensional space. Specifically, we initialize user embeddings as $\bm{u}_i \in \mathbb{R}^{d_1}$, item embeddings as $\bm{v}_j^s \in \mathbb{R}^{d_1}$ in the source domain and $\bm{v}_j^t \in \mathbb{R}^{d_1}$ in the target domain respectively, where $d_1$ is the embedding dimension.

\subsection{Personalized Preference Encoder}
This section introduces two parts: 1) a neural process (NP) encoder to generate the variational approximations over $\mathcal{T}_i$ and $\mathcal{C}_i$; 2) a normalizing flow (NF) encoder to convert the Gaussian (unimodal) distribution to a multimodal distribution to capture the user's personalized multi-interest preference.

\subsubsection{Neural Process Encoder}
NP regards each task $\mathcal{T}_i$ decoupled from the same stochastic process $f$. Specifically, we approximate $f$ using a random vector $\bm{z}_i$, which is generated by neural networks.
Thus, the neural process encoder aims to generate the variational prior $p(\bm{z}_i|\mathcal{C}_i)$ conditioned on support set $\mathcal{C}_i$ and variational posterior $q(\bm{z}_i|\mathcal{C}_i,\mathcal{Q}_i)$ with $\mathcal{T}_i$. 

Taking $\mathcal{C}_i$ as an example, for interactions in $\mathcal{C}_i$, we incorporate the $(\bm{u}_i,\bm{v}_j^s,y_{i,j}^s)$ to approximate $f$ using neural networks to produce corresponding representations $\bm{r}_{i,j} \in \mathbb{R}^{{|\mathcal{C}_i}|\times{d_2}}$, which can be formulated as:
\begin{equation}
\small
\begin{split}
\bm{r}_{i,j} = \mathrm{MLP} ([\bm{u}_i||\bm{v}_j^s||y_{i,j}^s];\phi),\quad (\bm{u}_i,\bm{v}_j^s,y_{i,j}^s)\in \mathcal{C}_i
\end{split}
\label{eq:np_concentration}
\end{equation}
where we use $\mathrm{MLP}(\cdot;\phi)$ parameterized by $\phi$ to fuse the features of $\mathcal{C}_i$. Besides, $[\cdot||\cdot]$ denotes the concatenation operation.

Afterword, to obtain representation $\bm{r}_i$, we aggregate $\bm{r}_{i,j}$ with an order-invariance operation as in NP studies~\cite{np,stochastic03}.
Therefore, taking $\mathcal{C}_i$ as an example, to derive $p(\bm{z}_i|\mathcal{C}_i)$, we generate $\bm{r}_i \in \mathbb{R}^{d_2}$ with a mean function for computational efficiency, the aggregation process is as follows:
\begin{equation}
\small
\begin{split}
\bm{r}_i = \frac{1}{|\mathcal{C}_i|}\sum_{j=1}^{|\mathcal{C}_i|}\bm{r}_{i,j},
\end{split}
\label{eq:np_aggregation}
\end{equation}

To obtain the random vector $\bm{z}_i$, we employ Gaussian distribution to generate $\bm{z}_i$ as in Variational inference studies~\cite{aevb}. Therefore, we samples $\bm{z}_i \sim \mathcal{N}(\bm{\mu}_i(\bm{r}_i),\bm{\sigma}_i(\bm{r}_i))$ as follows:
\begin{equation}
\small
\begin{split}
\bm{r}'_i &= \mathrm{ReLU}(\mathrm{MLP}(\bm{r}_i)),\\
\bm{\mu}_i(\bm{r}_i) &= \mathrm{MLP}(\bm{r}'_i),\\
\bm{\sigma}_i(\bm{r}_i) &= 0.1 + 0.9*\mathrm{Sigmoid}(\mathrm{MLP}(\bm{r}'_i)),
\end{split}
\label{np_distribution_1}
\end{equation}

Given that the random variable $\bm{z}_i$ is intractable during the training phase, we further adopt the reparameterization trick~\cite{aevb} to sample $\bm{z}_i$ from the probability distribution, which allows for smooth backpropagation of gradients:
\begin{equation}
\small
\begin{split}
\bm{z}_i = \bm{\mu}_i(\bm{r}_i) + \bm{\epsilon}\odot\bm{\sigma}_i(\bm{r}_i), \quad \bm{\epsilon} \sim \mathcal{N}(0,\bm{I}),
\end{split}
\label{np_distribution_2}
\end{equation}
where $\odot$ refers to element-wise multiplications. In this way, we can sample $\bm{z}_i$ from $p(\bm{z}_i|\mathcal{C}_i)$ and $q(\bm{z}_i|\mathcal{C}_i,\mathcal{Q}_i)$. Besides, formulating the users' interactions from both the source and target domains within a unified stochastic process enables the capture of users' personalized preference across domains.


\subsubsection{Normalizing Flow Encoder}
In general, the Gaussian distribution modeled by neural process encoder is a unimodal distribution. However, users typically exhibit multiple interests in real-world CDR scenarios. Towards this end, we are setting our sights on the normalizing flow. Taking $\mathcal{C}_i$ as an example, the key idea of normalizing flow is to convert the Gaussian (unimodal) distribution $p_0(\bm{z}_{i,0}|\mathcal{C}_i)$ to a multimodal distribution $p_K(\bm{z}_{i,K}|\mathcal{C}_i)$.



Specifically, $\bm{z}_{i,0}$ sampled through Eq.~(\ref{np_distribution_2}) is converted by $K$ steps of invertible and differentiable mappings $g_k$ to a multimodal distribution $p_K(\bm{z}_{i,K}|\mathcal{C}_i)$:
\begin{equation}
\small
\begin{split}
\bm{z}_{i,K} &= g_K \circ ... \circ g_2 \circ g_1(\bm{z}_{i,0}),\\
p_K(\bm{z}_{i,K}|\mathcal{C}_i) &= p_0(\bm{z}_{i,0}|\mathcal{C}_i)\prod_{k=1}^{K}\bigg|\mathrm{det}\frac{\partial g_k}{\partial \bm{z}_{i,k-1}}\bigg|^{-1},
\end{split}
\label{nf_distribution}
\end{equation}
where $\bm{z}_{i,K}$ contains more meaningful users' personalized multi-interest preference compared to $\bm{z}_{i,0}$. We provide a detailed analysis of the user's multi-interest preference captured by the normalizing flow in section \ref{Multimodal Distribution} and \ref{Case Study}.

\subsection{Common Preference Encoder}

This module consists of two parts: 1) a preference identity network to produce an elementary preference representation; 2) a preference pool to learn the common preference between different users in the support set $\mathcal{C}_i$.

\subsubsection{Preference Identity Network}

Given the support set $\mathcal{C}_i$ from source domain, the preference identify network tries to generate a elementary preference representation $\bm{e}_i \in \mathbb{R}^{d_3}$ by fusing the interactions in $\mathcal{C}_i$, which can be formulated as:
\begin{equation}
\small
\begin{split}
\bm{e}_{i,j} &= \mathrm{MLP} ([\bm{u}_i||\bm{v}_j^s||y_{i,j}^s];\theta),\quad (\bm{u}_i,\bm{v}_j^s,y_{i,j}^s)\in \mathcal{C}_i\\
\bm{e}_i &= \frac{1}{|\mathcal{C}_i|}\sum_{j=1}^{|\mathcal{C}_i|}\bm{e}_{i,j},
\end{split}
\label{ei}
\end{equation}
where we use $\mathrm{MLP}(\cdot;\theta)$ parameterized by $\theta$ to fuse the features of $\mathcal{C}_i$
and the elementary preference representation $\bm{e}_i$ is generated with the same operation in Eq.~(\ref{eq:np_aggregation}) for convenience.

\subsubsection{Preference Pool}
The preference pool $\mathcal{P} = [\bm{a}_1,...,\bm{a}_N]\in \mathbb{R}^{d_3\times N}$ serves as an additional resource which contains $N$ randomly initialized and trainable soft cluster centroids as $\bm{a}_n \in \mathbb{R}^{d_3}$. As suggested by~\cite{udeca,tanp}, we introduce an unsupervised algorithm aims to capture the common preference between different users, which operates through a two-step process. The first step involves generating soft cluster assignments between the elementary preference representation $\bm{e}_i$ and the preference pool. The second step focuses on updating the preference pool by integrating from high-confidence assignments with the help of an auxiliary distribution $\mathcal{D}$.


The Student’s t-distribution~\cite{t_sne} is used to measure the similarity between $\bm{e}_i$ and $\bm{a}_n$ as follows: 
\begin{equation}
\small
\begin{split}
\bm{c}_{in} = \frac{{(1 + {||\bm{e}_i-\bm{a}_n||}^2/\alpha)}^{-{\frac{\alpha+1}{2}}}}{\sum_{n'}{(1 + {||\bm{e}_i-\bm{a}_{n'}||}^2/\alpha)}^{-{\frac{\alpha+1}{2}}}},
\end{split}
\label{}
\end{equation}
where $\alpha$ are the degrees of freedom of the Student’s t-distribution. 
$\bm{c}_{in}$ can be understood as representing the probability of allocating $\bm{e}_i$ to $\bm{a}_n$, indicative of a soft assignment.
The final preference representation $\bm{h}_i \in \mathbb{R}^{d_3}$ is generated as:
\begin{equation}
\small
\begin{split}
\bm{h}_i = \bm{e}_i + \mathcal{P}\bm{c}_i^\top,
\end{split}
\label{eq:final_representation}
\end{equation}
where $\bm{c}_i\in \mathbb{R}^{N}$ is the similarity between $\bm{e}_i$ and $[\bm{a}_1,...,\bm{a}_N]$. Actually, the interactions in $\mathcal{C}_i$ represent the personal preference of user $\bm{u}_i$, and different users may have common preference which can be reflected by $\bm{c}_i$ through interacting with the preference pool $\mathcal{P}$. Consequently, the common preference between different users are comprised into the final preference representation $\bm{h}_i$ through Eq.~(\ref{eq:final_representation}). 

The soft assignments of all users in $\Omega^{tr}$ constitute $\mathcal{M} = [\bm{c}_1,...,\bm{c}_{|\Omega^{tr}|}]\in \mathbb{R}^{N\times |\Omega^{tr}|}$. To update the preference pool and refine the clusters, we propose an auxiliary distribution $\mathcal{D}$. Specifically, we define our objective as a KL divergence loss $\mathcal{L}_c$ between the soft assignments $\mathcal{M}$ and the auxiliary distribution $\mathcal{D}$ as follows:
\begin{equation}
\small
\begin{split}
\mathcal{D}_{in} &= \frac{(\mathcal{M}_{in})^2/\sum_i\mathcal{M}_{in}}{\sum_{n'}(\mathcal{M}_{in'})/\sum_i\mathcal{M}_{in'}},\\
\mathcal{L}_c &= \mathrm{K}\mathrm{L}(\mathcal{D}||\mathcal{M}) = \sum_i\sum_n\mathcal{D}_{in}\mathrm{log}\frac{\mathcal{D}_{in}}{\mathcal{M}_{in}},
\end{split}
\label{cluster_loss}
\end{equation}
Implementing $\mathcal{L}_c$ enhances the accuracy of predictions and prioritizes users assigned with high confidence. We provide a detailed analysis of the common preference between different users captured by NF-NPCDR in section~\ref{Visualization of preference pool}.


\subsection{Stochastic Adaptive Decoder}

To predict the rating score $\hat{y}_{i,j}^t$ with interactions in $\mathcal{Q}_i$, we have obtained the user's personalized multi-interest preference $\bm{z}_{i,K}$ and common preference $\bm{h}_i$ between different users. 
To enhance the integration of $\bm{z}_{i,K}$ and $\bm{h}_i$ for prediction, we introduce a stochastic adaptive decoder built upon the FiLM~\cite{film,ada,tanp}, capable of incorporating both the personalized and common preference for cold-start users.

For implementation, our stochastic adaptive decoder estimates the conditional likelihood $p(y_{i,j}^t|\bm{u}_i,\bm{v}_j^t,\bm{z}_{i,K})$. Initially, we generate two modulation parameters $\bm{\eta}_i$ and $\bm{\delta}_i$ over the representation $\bm{h}_i$, following which the stochastic adaptive decoder is formulated as follows:
\begin{equation}
\small
\begin{split}
\bm{\eta}_i^l&=\text{tanh}(\mathrm{MLP}(\bm{h}_i)),\\
\bm{\delta}_i^l&=\text{tanh}(\mathrm{MLP}(\bm{h}_i)),\\
\bm{w}^{0} &= [\bm{u}_i||\bm{v}_j^t||\bm{z}_{i,K}],\\
\bm{w}^{l+1} &= \text{ReLU}(\bm{\eta}_i^l\odot(\mathrm{MLP}(\bm{w}^{l}))+\bm{\delta}_i^l), \\
\end{split}
\label{decoder}
\end{equation}
where $\bm{w}^l$ denotes the input to the decoder, i.e., $l$-th layer. Specifically, $\bm{\eta}_i^l$ modulates the weight of $\bm{w}^l$, while $\bm{\delta}_i^l$ dominates the weight of $\bm{h}_i$.
Consequently, the stochastic adaptive decoder can effectively modulate the personalized and common preference for cold-start users, ultimately predicting the rating score $\hat{y}_{i,j}^t$ for the candidate item $\bm{v}_j^t$.

\subsection{Loss Function}

We define the loss function of our NF-NPCDR with three components: the log-likelihood to accomplish the desired CDR task, the KL divergence refined with the normalizing flow to impose regularization and KL divergence between soft assignments $\mathcal{M}$ and auxiliary distribution $\mathcal{D}$.

The desired log-likelihood for rating prediction in Eq.~(\ref{nf_np_loss}) is usually measured by Mean Square Error (MSE)~\cite{tanp, ptupcdr} as:
\begin{equation}
\small
\begin{split}
\mathcal{L}_{\mathrm{rec,i}} &= -E_{q_0(z_{i,0}|\mathcal{C}_i,\mathcal{Q}_i)}\log p(y_{i,1:|\mathcal{Q}_i|}^t|\bm{u}_i^o,\bm{v}_{1:|\mathcal{Q}_i|}^t,z_{i,K})\\
&\propto \frac{1}{|\mathcal{Q}_i|}\sum_{j=1}^{|\mathcal{Q}_i|}(y_{i,j}^t-\hat{y}_{i,j}^t)^2,
\end{split}
\label{rec_loss}
\end{equation}
where we use overlapping users' interactions in target domain $(\bm{u}_i^o,\bm{v}_{1:|\mathcal{Q}_i|}^t,y_{i,1:|\mathcal{Q}_i|}^t)$ to replace $(x_{1:|\mathcal{Q}|},y_{1:|\mathcal{Q}|})$. Moreover, we minimize the KL divergence in Eq.~(\ref{nf_np_loss}) by introducing the regularization term $\mathcal{L}_{\mathrm{KL,i}}$, which constrains the approximate posterior of $\mathcal{Q}_i$ toward the conditional prior of $\mathcal{C}_i$:
\begin{equation}
\small
\begin{split}
\mathcal{L}_{\mathrm{KL,i}} &= E_{q_0(z_{i,0}|\mathcal{C}_i,\mathcal{Q}_i)}\bigg[\log q_0(\bm{z}_{i,0}|\mathcal{C}_i,\mathcal{Q}_i) \\
&\;\;\;\;\;\;\;\;\;\;\;\;\; - \sum_{k=1}^{K}\bigg|\mathrm{det}\frac{\partial g_k}{\partial \bm{z}_{i,k-1}}\bigg| - \log p(\bm{z}_{i,K}|\mathcal{C}_i)\bigg],
\end{split}
\label{kl_loss}
\end{equation}
In conclusion, the total optimization function is formulated as:
\begin{equation}
\small
\begin{split}
\mathcal{L} = \frac{1}{|\Omega^{tr}|}\sum_{i=1}^{|\Omega^{tr}|}(\mathcal{L}_{\mathrm{rec,i}} + \mathcal{L}_{\mathrm{KL,i}}) + \lambda \mathcal{L}_c,
\end{split}
\label{total_loss}
\end{equation}
where $\lambda$ is a hyper-parameter (ranging from 0 to 1) to balance $\mathcal{L}_c$ from Eq.~(\ref{cluster_loss}). 
The training algorithm of NF-NPCDR is shown in Algorithm~\ref{alg1}.

\begin{algorithm}[tb]
\caption{for NF-NPCDR in the training phase}
\footnotesize
\label{alg1}
\textbf{Input}: Training overlapping users set $\mathcal{U}^o$; Items set $\mathcal{V}^s$ and $\mathcal{V}^t$; User-item rating matrix $\mathcal{Y}^s$ and $\mathcal{Y}^t$; Training tasks set $\Omega^{tr}$. \\
\textbf{Output}: Model parameters $\Theta$.
\begin{algorithmic}[1] 
\STATE Initialize all model parameters. \\
\STATE \textbf{while} not convergence \textbf{do} \\
\STATE\quad \textbf{for} $\mathcal{T}_i \in \Omega^{tr}$ \textbf{do} \\ 
\STATE\quad\quad Construct $\mathcal{C}_i$ and $\mathcal{Q}_i$ from $\mathcal{T}_i$. \\
\STATE\quad\quad Apply NP to generate $q_0(\bm{z}_{i,0}|\mathcal{C}_i,\mathcal{Q}_i)$ in Eq.~(\ref{np_distribution_1})-(\ref{np_distribution_2}). \\
\STATE\quad\quad Sample a $\bm{z}_{i,0}$ from $q_0(\bm{z}_{i,0}|\mathcal{C}_i,\mathcal{Q}_i)$. \\
\STATE\quad\quad Apply NF to generate $\bm{z}_{i,K}$ in Eq.~(\ref{nf_distribution}).\\
\STATE\quad\quad Generate $\bm{h}_i$ in Eq.~(\ref{ei})-(\ref{eq:final_representation}).\\
\STATE\quad\quad Predictions on $\mathcal{Q}_i$ with $\bm{z}_{i,K}$ and $\bm{h}_i$ in Eq.~(\ref{decoder}). \\
\STATE\quad\quad Generate $p(\bm{z}_i|\mathcal{C}_i)$ using $\mathcal{C}_i$ in Eq.~(\ref{np_distribution_1}) and Eq.~(\ref{np_distribution_2}). \\
\STATE\quad\quad Calculate $\mathcal{L}_{\mathrm{rec,i}}$ in Eq.~(\ref{rec_loss}) and $\mathcal{L}_{\mathrm{KL,i}}$ in Eq.~(\ref{kl_loss}).\\
\STATE\quad \textbf{end for}\\
\STATE\quad Calculate $\mathcal{L}_c$ in Eq.~(\ref{cluster_loss}).\\
\STATE\quad Use the overall loss $\mathcal{L}$ in Eq.~(\ref{total_loss}) to optimize $\Theta$.\\
\STATE\textbf{end while}
\end{algorithmic}
\end{algorithm}

Note that in the testing phase, 
given a cold-start user $u_i^{s\setminus o}\in \mathcal{U}^{s\setminus o}$, we first enhance the neural process with the normalizing flow to generate $p_K(\bm{z}_{i,K}|\mathcal{C}_i)$ from $\mathcal{C}_i$. Then, we sample a random variable $\bm{z}_{i,K}$ from $p_K(\bm{z}_{i,K}|\mathcal{C}_i)$. And we generate $\bm{h}_i$ from the common preference encoder. Last, we take $\bm{z}_{i,K}$, $\bm{h}_i$ and $(\bm{u}_i^{s\setminus o},\bm{v}_j^t)$ in $\mathcal{Q}_i$ as inputs to the stochastic adaptive decoder to predict $\hat{y}_{i,j}^t$ for $\mathcal{Q}_i$.
The testing algorithm of NF-NPCDR is shown in Algorithm~\ref{alg2}.

\begin{algorithm}[tb]
\caption{for NF-NPCDR in the testing phase}
\footnotesize
\label{alg2}
\textbf{Input}: Testing cold-start users set $\mathcal{U}^{s\setminus o}$; Items set $\mathcal{V}^s$ and $\mathcal{V}^t$; User-item rating matrix $\mathcal{Y}^s$; Testing tasks set $\Omega^{te}$; Model parameters $\Theta$. \\
\textbf{Output}: Prediction rating score $\hat{y}_{i,j}^t$.
\begin{algorithmic}[1] 
\STATE \textbf{for} $\mathcal{T}_i \in \Omega^{te}$ \textbf{do} \\ 
\STATE\quad Construct $\mathcal{C}_i$ and $\mathcal{Q}_i$ from $\mathcal{T}_i$. \\
\STATE\quad Apply NP to generate $p_0(\bm{z}_{i,0}|\mathcal{C}_i)$ in Eq.~(\ref{np_distribution_1})-(\ref{np_distribution_2}). \\
\STATE\quad Sample a $\bm{z}_{i,0}$ from $p_0(\bm{z}_{i,0}|\mathcal{C}_i)$. \\
\STATE\quad Apply NF to generate $\bm{z}_{i,K}$ in Eq.~(\ref{nf_distribution}).\\
\STATE\quad Generate $\bm{h}_i$ in Eq.~(\ref{ei})-(\ref{eq:final_representation}).\\
\STATE\quad Predictions on $\mathcal{Q}_i$ with $\bm{z}_{i,K}$ and $\bm{h}_i$ in Eq.~(\ref{decoder}). \\
\STATE \textbf{end for}\\
\end{algorithmic}
\end{algorithm}

\subsection{Complexity}

In NF-NPCDR, all modules are parameterized by MLP, including the personalized preference encoder, common preference encoder and stochastic adaptive decoder. Therefore, our model keeps an efficient architecture for training and inference. Specifically, the complexity of each module can be approximated as $\mathcal{O}(l{d^3})$, where $l$ represents the number of MLP layers and $d$ is the hidden size of each layer. The complexity of computing the Jacobian determinant of the normalizing flow in Eq.~(\ref{nf_distribution}) is $\mathcal{O}(Kd)$, where $K$ is the flow-steps. The calculation of final preference representation $\bm{h}_i$ costs $\mathcal{O}(N{d^2})$, where $N$ represents the number of soft cluster centroids in preference pool. In addition, the FiLM requires only two weight metrics per layer in stochastic adaptive decoder, which is computationally efficient. Overall, NF-NPCDR is a lightweight and efficient framework.

\section{Experiments}\label{experiments}

In this section, we conduct extensive experiments on three real-world CDR scenarios to answer the following research questions (RQs):
(1) \textbf{RQ1}: Compared with other state-of-the-art CDR models, does our model achieve the significant performance?
(2) \textbf{RQ2}: Does the normalized flow proposed in our model really capture user's personalized multi-interest preference?
(3) \textbf{RQ3}: Does the preference pool proposed in our model really capture the common preference between different users?
(4) \textbf{RQ4}: Can directly leveraging the neural process to bridge the source and target domains really achieve better experimental performance?
(5) \textbf{RQ5}: What is the effect of different normalizing flows on NF-NPCDR? Is there a trade-off between experimental performance and computational speed of normalizing flow?
(6) \textbf{RQ6}: How does the computational cost of NF-NPCDR?

\subsection{Datasets and Metrics}

Following most existing methods~\cite{sscdr,cdml,tmcdr,ptupcdr,cdrnp}, we evaluate the performance of our NF-NPCDR against other baselines on two real-world datasets (\textit{i.e.}, \textbf{Amazon}\footnote{\url{http://jmcauley.ucsd.edu/data/amazon/}} and \textbf{Douban}\footnote{\url{https://www.douban.com/}}) as follows:
\begin{itemize}
    \item The Amazon dataset collects user review data from \textit{amazon.com}, one of the largest e-commerce sites in the world. The collected data spans from May 1996 to July 2014.
    \item The Douban dataset is collected from \textit{douban.com}, a popular service in China that provides ratings in various categories.
\end{itemize}
On the one hand, there have been many works~\cite{cmif,epimi,miss,wmlmmi,imsr,re4,comorec} using Amazon/Douban dataset to perform multi-interest recommendation. On the other hand, many works~\cite{cdml,sscdr,cdrnp,emcdr,catn,ptupcdr,tmcdr} also adopt Amazon/Douban dataset to alleviate the cold-start issue. Therefore, the Amazon and Douban datasets are widely used in the field of recommendation systems to explore multi-interest cold-start case. In this paper, we choose Amazon and Douban datasets to verify the effectiveness of our model in capturing the multi-interest preference of cold-start users.

The Amazon dataset consists of 24 various item domains. 
Since users tend to have similar preferences in relevant domains, following previous methods~\cite{cdrnp,remit,ptupcdr},
we select three relevant domains from Amazon, including Book (named "Books" in Amazon), Movie (named "Movie and TV" in Amazon) and Music (named "CDs and Vinyl" in Amazon) to form 
$\textit{Scenario 1}$ : Amazon-Movie $\to$ Amazon-Music,
$\textit{Scenario 2}$ : Amazon-Book $\to$ Amazon-Movie, and $\textit{Scenario 3}$ : Amazon-Book $\to$ Amazon-Music. 
Similarly, we select three relevant domains from Douban, including Movie, Book and Music to form \textit{scenario 4}: Douban-Movie $\to$ Douban-Book and \textit{scenario 5}: Douban-Movie $\to$ Douban-Music.
Both datasets contain rating scores from 1 to 5, reflecting user's preference for specific items. For each CDR scenario, we first filter out users and items with fewer than 5 interactions. Then, we further filter out the items with user rating scores less than 4 in the source domain to eliminate noise information. Moreover, we create user interaction sequences in the source/target domain based on the sequential timestamps.
Unlike many existing works~\cite{dfrcdr,ncf,cml,cdtf,BPR,sscdr} that only utilize a subset of the dataset for evaluation, we employ the entire dataset to better simulate real-world application scenarios.
Table~\ref{sec_exp_tab:dataset} summarizes the details of five CDR scenarios.

Following previous works~\cite{emcdr,catn,ptupcdr}, We adopt two widely used metrics, Mean Absolute Error (MAE) and Root Mean Square Error (RMSE) 
for performance comparison.

\begin{table}[t]
\centering
\scriptsize
\caption{Statistics of Five CDR scenarios on Amazon and Douban datasets (\#Overlap denotes the number of overlapping users).}
\setlength{\tabcolsep}{4.5pt}
\begin{tabular*}{0.47 \textwidth}{@{\extracolsep{\fill}}@{}l|lrrrr@{}}
\toprule
{\bf Scenarios}  & \bf Domain  &  {\bf \#Users}  &  {\bf \#Overlap}  &  {\bf \#Items}  &  {\bf \#Ratings}\\
\midrule
\multirow{2}{*}{\bf Scenario 1}  & {$\mathcal{D}^s$ : {Movie}}  &123,960   &\multirow{2}{*}{18,031}   & 
 {50,052}  &1,697,533\\ 
& {$\mathcal{D}^t$ : {Music}}  &75,258  &  &  {64,443}  &1,097,592\\
\midrule
\multirow{2}{*}{\bf Scenario 2}  & {$\mathcal{D}^s$ : {Book}}  &603,668   &\multirow{2}{*}{37,388}   &{367,982}  &8,898,041\\ 
&{$\mathcal{D}^t$ : {Movie}}   &123,960  &  &  {50,052}  &1,697,533\\ 
\midrule
\multirow{2}{*}{\bf Scenario 3}  & {$\mathcal{D}^s$ : {Book}}  &603,668   &\multirow{2}{*}{16,738}   &{367,982}  &8,898,041\\ 
&{$\mathcal{D}^t$ : {Music}}   &75,258  &  &  {64,443}  &1,097,592\\
\midrule
\multirow{2}{*}{\bf Scenario 4}  & {$\mathcal{D}^s$ : {Movie}}  &2,712   &\multirow{2}{*}{2,209}   & 
 {34,893}  &1,278,401 \\
& {$\mathcal{D}^t$ : {Book}}  &2,212  &  &  {95,872}  &227,251 \\
\midrule
\multirow{2}{*}{\bf Scenario 5}  & {$\mathcal{D}^s$ : {Movie}}  &2,712   &\multirow{2}{*}{1,815}   &{34,893}  &1,278,401 \\ 
&{$\mathcal{D}^t$ : {Music}}   &1,820  &  &  {79,878}  &17,9847 \\
\bottomrule
\end{tabular*}
\vspace{-0.5cm}
\label{sec_exp_tab:dataset}
\end{table}



\subsection{Baselines}

In our experiments, we compare our NF-NPCDR with the following state-of-the-art baselines: 
(1) \textbf{TGT}~\cite{tgt} denotes the target matrix factorization model, which concentrates solely on interactions within the target domain, ignoring those from the source domain.
(2) \textbf{CMF}~\cite{cmf} is a collective matrix factorization method, involves decomposing multiple matrices simultaneously and sharing parameters among users across domains.
(3) \textbf{EMCDR}~\cite{emcdr} is the first work that proposes an Embedding and Mapping paradigm to cold-start users. It generates user/item representations in both domains, then utilizes a mapping function to align the users' representations.
(4) \textbf{CATN}~\cite{catn} proposes to facilitate the transfer of user preferences at the aspect level and employs an attention mechanism to learn the aspect correlations across domains.
(5) \textbf{DCDCSR}~\cite{dcdcsr} enhances the EMCDR paradigm by introducing benchmark factors. This method considers the varying degrees of rating sparsity for single users or items across different domains and systems.
(6) \textbf{SSCDR}~\cite{sscdr} introduces a CDR framework that utilizes a semi-supervised mapping approach. It models the cross-domain relationship with the help of mapping function.
(7) \textbf{LACDR}~\cite{lacdr} aims to reduce the training difficulty of the mapping function, which maintains high expressiveness and facilitates effective optimization.
(8) \textbf{RecGURU}~\cite{recguru} proposes an adversarial learning method to achieve cross-domain collaboration in user representations.
(9) \textbf{PTUPCDR}~\cite{ptupcdr} uses a meta-learning-based framework to bridge the source and target domains with personalized mapping functions for each user.
(10) \textbf{REMIT}~\cite{remit} constructs a heterogeneous information network to generate user's multiple deterministic interest embeddings across domains. 
(11) \textbf{CDRNP}~\cite{cdrnp} uses the attention mechanism to bridge the source and target domains, then introduces the neural process to capture user preference correlations within domains.

In contrast, we exploit neural process to explicitly transfer users' personalized preference across domains, and further adopt normalizing flow to fully model the user's personalized multi-interest preference.

\subsection{Implementation Details} 

The experiments are conducted in a server with Intel(R) Xeon(R) Silver 4110 CPU @ 2.10GHz, and Tesla T4 (16G) GPU. The above methods are implemented in Pytorch 1.9.0 with python 3.6.8.
In our experiments, we adjust the input to models and evaluation metrics of all baselines based on their official implementations to align with our experimental settings. To ensure the fairness of comparison, we adopt the identical value for the common hyper-parameters, including: the dimensions of the user/item representations are set to 10, the learning rate is set to 0.01, the batch size is set to 128, the number of MLP layers is set to 3 and the hidden size of each fully connected layer is set to 64. For the specific hyper-parameters in the baselines, we use the values reported in their original literature. For our NF-NPCDR, we select the Planar flow~\cite{planarflow} to serve as the normalizing flow in Eq.~(\ref{nf_distribution}). The number of the transformation steps $K$ in Eq.~(\ref{nf_distribution}) is chosen from $\left\{{2, 4, 6, 8, 10, 12, 14, 16}\right\}$. 
The number $N$ of soft cluster centroids in the preference pool $\mathcal{P}$ is chosen from $\left\{{10, 20, 30, 40, 50, 60}\right\}$. The hyper-parameter $\lambda$ in Eq.~(\ref{total_loss}) is chosen from $\left\{{0.1, 0.2, 0.3, 0.4, 0.5, 0.6, 0.7, 0.8, 0.9, 1.0}\right\}$, the length of the support set $\mathcal{C}_i$ is chosen from $\left\{{5, 10, 15, 20, 25}\right\}$. We consider sequential timestamps in our experiments to prevent information leakage. For each CDR scenario, the best hyper-parameters are tuned by grid search according to Mean Absolute Error (MAE) over five random runs.

\begin{table*}[t]
\scriptsize
\centering
\caption{Overall performance comparison on Amazon dataset.}
\label{mainexperiment}
\setlength\tabcolsep{2.0pt}
\begin{tabular*}{1 \textwidth}
{@{\extracolsep{\fill}}@{}lcccccc|cccccc|cccccc@{}}
\toprule
&
\multicolumn{6}{c}{\bf Amazon-Movie $\to$ Amazon-Music} & \multicolumn{6}{c}{\bf Amazon-Book $\to$ Amazon-Movie} & \multicolumn{6}{c}{\bf Amazon-Book $\to$ Amazon-Music}    \\
\cmidrule(r){2-7}\cmidrule(r){8-13}\cmidrule(r){14-19} \bf Methods &
\multicolumn{2}{c}{ \bf{20\%}} & \multicolumn{2}{c}{ \bf{50\%}} & \multicolumn{2}{c}{ \bf{80\%}} &  \multicolumn{2}{c}{ \bf{20\%}} & \multicolumn{2}{c}{ \bf{50\%}} & \multicolumn{2}{c}{ \bf{80\%}} &  \multicolumn{2}{c}{ \bf{20\%}} & \multicolumn{2}{c}{ \bf{50\%}} & \multicolumn{2}{c}{ \bf{80\%}} \\
\cmidrule(r){2-3}\cmidrule(r){4-5}\cmidrule(r){6-7}
\cmidrule(r){8-9}\cmidrule(r){10-11}\cmidrule(r){12-13}
\cmidrule(r){14-15}\cmidrule(r){16-17}\cmidrule(r){18-19}&
\bf{MAE} & \bf{RMSE} & \bf{MAE} & \bf{RMSE} & \bf{MAE} & \bf{RMSE} & \bf{MAE} & \bf{RMSE} & \bf{MAE} & \bf{RMSE} & \bf{MAE} & \bf{RMSE} & \bf{MAE} & \bf{RMSE} & \bf{MAE} & \bf{RMSE} & \bf{MAE} & \bf{RMSE} \\
\midrule
TGT~\cite{tgt}  &   4.4803   &   5.1580   &   4.4989   &   5.1736   &   4.5020   &   5.1891   & 
 4.1831   &   4.7536   &   4.2288   &   4.7920   &   4.2123   &   4.8149   &   4.4873   &   5.1672   &   4.5073   &   5.1727   &   4.5204   &   5.2308  \\

CMF~\cite{cmf}    &  1.5209 &   2.0158    &   1.6893    &    2.2271 &   2.4186  &   3.0936 & 
 1.3632   &   1.7918 &  1.5813   &    2.0886  &    2.1577    &    2.6777  &  1.8284  &  2.3829  &  2.1282  &  2.7275  &  3.0130  &  3.6948  \\
DCDCSR~\cite{dcdcsr} &  1.4918 &   1.9210    &   1.8144    &    2.3439 &   2.7194  &   3.3065 & 
 1.3971   &   1.7346 &  1.6731   &    2.0551  &    2.3618    &    2.7702  &  1.8411  &  2.2955  &  2.1736  &  2.6771  &  3.1405  &  3.5842  \\
SSCDR~\cite{sscdr}   &    1.3017 &   1.6579    &   1.3762    &    1.7477 &   1.5046  &   1.9229 & 
 1.2390   &   1.6526 &  1.2137   &    1.5602  &    1.3172    &    1.7024  &  1.5414  &  1.9283  &  1.4739  &  1.8441  &   1.6414  &  2.1403  \\
EMCDR~\cite{emcdr}   &  1.2350 &   1.5515 &  1.3277 &  1.6644 &  1.5008  &  1.8771  &
 1.1162 &  1.4120   &  1.1832 &  1.4981 &  1.3156 &  1.6433  &  1.3524  &  1.6737  &  1.4723  &  1.8000  &  1.7191  &  2.1119  \\
CATN~\cite{catn}  &  1.2671 &   1.6468 &  1.4890 &  1.9205 &  1.8182  &  2.2991  &  1.1249 &  1.4548   &  1.1598 &  1.4826 &  1.2672 &  1.6280  &  1.3924  &  1.7399  &  1.6023  &  2.0665  &  1.9571  &  2.5623 \\
LACDR~\cite{lacdr}  &  1.1295  &  1.4358  &   1.3502  &  1.7510  &  1.6886  &  2.2238  &  0.9681 &  1.2311   &  1.0077 &  1.3051 &  1.1151 &  1.4660  &  1.1945  &  1.5771  &  1.3925  &  1.8644  &  1.7107  &  2.2468  \\
RecGURU~\cite{recguru}  &  1.2320  &  1.5545   &   1.3696   &  1.6640  &  1.7154  &  2.2160  &  1.0404 &  1.2598   &  1.2434 &  1.4921 &  1.2243 &  1.5801  &  1.4467  &  1.7188  &  1.7496  &  2.0766  &  1.8535  &  2.3401 \\
PTUPCDR~\cite{ptupcdr}  &   1.1504 &   1.5195 &  1.2804 &  1.6380 &  1.4049  &  1.8234  &   0.9970 &  1.3317   &  1.0894 &  1.4395 &  1.1999 &  1.5916  &   1.2286  &  1.6085  &  1.3764  &  1.7447  &  1.5784  &  2.0510 \\
REMIT~\cite{remit} &  0.9393  &  1.2709   &   1.0437   &  1.4580  &  1.2181  &  1.6601  &   \underline{0.8759} &  1.1650   &  0.9172 &  1.2379 &  1.0055 &  1.3772  &  1.3749  &  1.9940  &  1.4401  &  2.0495  &  1.6396  &  2.2653 \\
CDRNP~\cite{cdrnp}  &  \underline{0.7974} &  \underline{1.0638}   &  \underline{0.7969} &  \underline{1.0589} &  \underline{0.8280} &  \underline{1.0758}  &  0.8846 &  \underline{1.1327}   &  \underline{0.8946} &  \underline{1.1450} &  \underline{0.8970} &  \underline{1.1576}  &  \underline{0.7453}  &  \underline{0.9914}  &  \underline{0.7629}  &  \underline{1.0111}  &  \underline{0.7787}  &  \underline{1.0373} \\
\midrule
NF-NPCDR &  \textbf{0.4348}  &  \textbf{0.8146}   &   \textbf{0.4479}   &  \textbf{0.8310}  &  \textbf{0.4598}  &  \textbf{0.8399}  &  \textbf{0.4490}  &  \textbf{0.8511}  &  \textbf{0.4445}  &  \textbf{0.8643}  &  \textbf{0.4746}  &  \textbf{0.8828}  &  \textbf{0.3894}  &  \textbf{0.7632}  &  \textbf{0.4027}  &  \textbf{0.7706}  &  \textbf{0.4170}  &  \textbf{0.7748} \\
Improve & {45.47\%}  &  {23.43\%}  &  {43.79\%}  &  {21.52\%}  &  {44.47\%}  &  {21.93\%} & {48.74\%}  &  {24.86\%}  &  {50.31\%}  &  {24.52\%}  &  {47.09\%}  &  {23.74\%}  & {47.75\%}  &  {23.02\%}  &  {47.21\%}  &  {23.79\%}  &  {46.45\%}  &  {25.31\%}\\
\bottomrule
\end{tabular*}
\end{table*}

\begin{table*}[t]
\scriptsize
\centering
\caption{Overall performance comparison on Douban dataset.}
\label{main_douban}
\setlength\tabcolsep{2.0pt}
\begin{tabular*}{0.8 \textwidth}
{@{\extracolsep{\fill}}@{}lcccccc|cccccc@{}}
\toprule
&
\multicolumn{6}{c}{\bf Douban-Movie $\to$ Douban-Book} & \multicolumn{6}{c}{\bf Douban-Movie $\to$ Douban-Music}    \\
\cmidrule(r){2-7}\cmidrule(r){8-13} \bf Methods &
\multicolumn{2}{c}{ \bf{20\%}} & \multicolumn{2}{c}{ \bf{50\%}} & \multicolumn{2}{c}{ \bf{80\%}} &  \multicolumn{2}{c}{ \bf{20\%}} & \multicolumn{2}{c}{ \bf{50\%}} & \multicolumn{2}{c}{ \bf{80\%}}  \\
\cmidrule(r){2-3}\cmidrule(r){4-5}\cmidrule(r){6-7}
\cmidrule(r){8-9}\cmidrule(r){10-11}\cmidrule(r){12-13}&
\bf{MAE} & \bf{RMSE} & \bf{MAE} & \bf{RMSE} & \bf{MAE} & \bf{RMSE} & \bf{MAE} & \bf{RMSE} & \bf{MAE} & \bf{RMSE} & \bf{MAE} & \bf{RMSE} \\
\midrule
TGT~\cite{tgt}  &  4.3608    &   5.1976   &   4.2722   &   5.0742   &   4.2999   &   5.1121   & 4.3225
    &   5.1000   &   4.3590   &   5.1262   &   4.3877   &   5.1655   \\

CMF~\cite{cmf}    &   2.3455   &   3.1429   &   2.5414   &   3.3250   &  3.0902    &   3.8110   & 2.5785
    &   3.3743   &   2.7830   &   3.5396   &   3.2192   & 3.9194 \\
DCDCSR~\cite{dcdcsr} &  2.2801    &   2.7748   &  2.5499    &   3.0822   & 3.3885     &   3.6213   & 2.1690
    &   2.5646   &   2.7990   &   3.1982   &   3.6585   & 3.8855 \\
SSCDR~\cite{sscdr}   &   2.2723   &   2.8064   &   2.4033   &   2.9383   & 3.2540     &   3.4689   & 2.2496
    &   2.5602   &   3.0579   &   2.9516   &   3.3226   & 3.4409 \\
EMCDR~\cite{emcdr}   &  2.4435    &   3.0077   &   2.6365   &   3.0877   & 3.4904     &   3.7434   & 2.3045
    &   2.8589   &   2.6926   &   3.1398   &   3.5168   & 3.7645 \\
CATN~\cite{catn}  &   2.3285   &   2.9363   &   2.4721   &   2.9764   & 3.2250     &   3.4573   & 2.2685
    &   2.7831   &   2.8840   &   3.0276   &   3.3078   & 3.4930 \\
LACDR~\cite{lacdr}  &   2.2032   &  2.7317    &   2.8873   &   3.4064   & 2.9873     &   3.2064   & 2.8020
    &   3.3021   &   3.0005   &   3.3544   &   3.4071   & 3.6752 \\
RecGURU~\cite{recguru}  &   2.2579   &   2.8931   &   2.6720   &   3.1351   &   3.0704   &   3.2916   & 2.2943
    &   2.6009   &   2.7163   &   2.9245   &   3.2880   & 3.4061 \\
PTUPCDR~\cite{ptupcdr}  &   1.6914   &   2.3176   &   1.9729   &   2.5599   &   2.8191   &   3.2046   & 1.9116
    &   2.5670   &   2.1192   &   2.7271   &   2.8990   & 3.3057 \\
REMIT~\cite{remit} &   1.4628   &   2.0757   &   1.8023   &   2.3849   &  2.7584    &   2.9835   & 1.6657
    &   2.4962   &   2.0270   &   2.5538   &   2.7376   & 3.1890 \\
CDRNP~\cite{cdrnp}  &   \underline{0.7350}   &   \underline{0.9315}   &   \underline{0.7363}   &   \underline{0.9470}   &  \underline{0.7382}    &   \underline{0.9461}   & \underline{0.6971}
    &   \underline{0.8937}   &   \underline{0.7376}   &   \underline{0.9278}   &   \underline{0.7505}   & \underline{0.9428} \\
\midrule
NF-NPCDR &   \textbf{0.4236}   &   \textbf{0.4861}   &   \textbf{0.4148}   &   \textbf{0.4703}   &   \textbf{0.4299}   & \textbf{0.4839}     & \textbf{0.4339}
    &  \textbf{0.5012}    &   \textbf{0.4233}   &   \textbf{0.4813}   &   \textbf{0.4581}   & \textbf{0.4975} \\
Improve &   42.37\%   &   47.82\%   &   43.66\%   &   50.34\%   &   41.76\%   &   48.85\%   & 37.76\%
    &   43.92\%   &   42.61\%   &   48.12\%   &   38.96\%   & 47.23\% \\
\bottomrule
\end{tabular*}
\vspace{-0.3cm}
\end{table*}

Following previous methods~\cite{emcdr,ptupcdr,cdrnp}, we randomly remove all the ratings from a subset of overlapping users in the target domain, treating these users as testing (cold-start) users, while considering the remaining overlapping users as training users. Suggested by previous methods~\cite{cdrnp,remit,ptupcdr}, we set 20$\%$, 50$\%$, and 80$\%$ of the overlapping users as the proportion of testing (cold-start) users, denoted as $\alpha$. In fact, different overlapping user ratio (\textit{i.e.}, $1-\alpha$) represents different situations, \textit{e.g.}, $\alpha$ = 20\% represents most of users are overlapped, while $\alpha$ = 80\% means only few users are overlapped.
Following previous meta-learning recommenders~\cite{mlcdr,mamo,melu}, we focus on predicting the ratings of each user for the candidate items in the query set $\mathcal{Q}_i$.

\subsection{Performance Comparisons (RQ1)}

Table~\ref{mainexperiment} and \ref{main_douban} show the comparison results on five CDR scenarios according to MAE and RMSE. From the experimental results, we observe that NF-NPCDR shows consistent prediction improvement compared with the single-domain baselines (i.e., TGT). This observation indicates that devising information transfer strategies across domain is effective to improve the recommendation performance in target domain. We also observe that NF-NPCDR consistently outperforms the cross-domain baselines in all five CDR scenarios, indicating the effectiveness of our model. Specifically, with different $\alpha$ value settings at 20$\%$, 50$\%$ and 80$\%$, NF-NPCDR surpasses the current state-of-the-art method (i.e., CDRNP) in terms of MAE by an average of 47.32$\%$, 47.10$\%$ and 46.00$\%$ in Amazon dataset, and 40.07$\%$, 43.14$\%$ and 40.36$\%$ in Douban dataset, respectively. This observation validates that capturing the user's personalized multi-interest preference and common preference between different users are effective for CDR. 

We perform a further introspection of our results and note that the multimodal distribution modeled by NF-NPCDR can capture richer and more meaningful information for CDR. Compared to the EMCDR-based CDR methods, our NF-NPCDR achieves better performance, which indicates that a multimodal distribution can capture the user's multi-interest preference. Meanwhile, we find that compared with the SOTA meta-learning method (i.e., CDRNP), NF-NPCDR exhibits exceptional performance and robustness, even in scenarios where the training data (overlapping users) is limited. Specifically, for Amazon-Movie $\to$ Amazon-Music,
even if the $\alpha$ value of the SOTA meta-learning method is set to 20\% and the $\alpha$ value of our method is set to 80\%, we can still obtain an experimental performance improvement in MAE of 42.34\%. In other words, our method uses only 20\% training data can significantly outperform the SOTA method with 80\% training data. This indicates that our normalizing flow enhanced neural process framework is capable of adapting to the CDR task with only a limited number of overlapping users.

\begin{table}[t]
\centering
\scriptsize
\caption{Ablation Study on five CDR scenarios.}
\label{ablation}
\setlength\tabcolsep{2pt}
\begin{tabular*}{0.47 \textwidth}{@{\extracolsep{\fill}}@{}lc|ccccc@{}}
\toprule
\bf \text {Variants} & \bf \text {Metric} & \bf \text {Scenario1}  & \bf \text {Scenario2} & \bf \text {Scenario3} & \bf \text {Scenario4} & \bf \text {Scenario5}  \\
\midrule
NF-NPCDR & \text {MAE} &   0.4348 &  0.4490 &  0.3894 &  0.4236  &  0.4339 \\
(NF+PO+FM)&\text {RMSE} &   0.8146 &  0.8511 &  0.7632 &  0.4861  &  0.5012 \\
\midrule
\multirow{2}{*}{\textit{w}/\textit{o} NF}   &   \text {MAE}  &  0.4660   &  0.4606 &   0.4132 &  0.4585   &  0.4671  \\
&\text {RMSE} &  0.8320   &  0.8647 &   0.7761 &  0.5082   &  0.5209\\
\midrule
\multirow{2}{*}{\textit{w}/\textit{o} PO} & \text {MAE} &   0.4722   &  0.4689 &  0.4127 &   0.4510   &   0.4574    \\
&\text {RMSE} & 0.8404   &  0.8605 &  0.7781 &  0.5074  &  0.5188 \\
\midrule
\multirow{2}{*}{\textit{w}/\textit{o} FM} & \text {MAE} &   0.4468   &  0.4561 &  0.4009 &   0.4476   &  0.4513    \\
&\text {RMSE} & 0.8224   &  0.8592 &  0.7697 &  0.4930  &  0.5119 \\
\midrule
\multirow{2}{*}{\textit{w}/\textit{o} NF, PO, FM} & \text {MAE} &   0.4922   &  0.4865 &  0.4304 &   0.4770   &   0.4899    \\
&\text {RMSE} & 0.8512   &  0.8735 &  0.7842 &  0.5215  &  0.5473 \\

\bottomrule
\end{tabular*}
\vspace{-0.3cm}
\end{table}

\subsection{Ablation Study}

In this section, we conduct ablation studies to analyze the effectiveness of each component in NF-NPCDR. We conduct four variants of NF-NPCDR in Table~\ref{ablation}. Specifically, the \textit{w}/\textit{o} NF is the variant without normalizing flow encoder, the \textit{w}/\textit{o} PO is the variant without preference pool, the \textit{w}/\textit{o} FM is the variant without FiLM, and the \textit{w}/\textit{o} NF, PO, FM is the variant without the three components mentioned above. We set the $\alpha$ value for all CDR scenarios to 20{\%}. 

Table \ref{ablation} presents the results. We find that MAE of our model drops by 6.69\%, 2.52\%, 5.76\%, 7.61\% and 7.11\% after the normalizing flow encoder is removed in five CDR scenarios. This demonstrates that the importance of normalizing flow encoder to further capture the user's personalized multi-interest preference. 
We also observe that MAE of our model drops by 7.92\%, 4.24\%, 5.65\%, 6.08\% and 5.14\% after the preference pool is removed. This indicates that the common preference between different users captured by the preference pool is effective for CDR. 
Meanwhile, MAE of our model drops 2.69\%, 1.56\%, 2.87\%, 5.36\% and 3.86\% after the FiLM is removed, which demonstrates that the FiLM can indeed incorporate both the personalized and common preference for cold-start users, adaptively modulate both preference for better recommendation. 
Lastly, compared to \textit{w}/\textit{o} NF, PO, FM variant, the normalizing flow, preference pool and FiLM can contribute a total of 11.66\%, 7.71\%, 9.53\%, 11.19\% and 11.43\% increase in MAE to our model. Thus, we should combine them to achieve the best performance and generalizability.



\begin{figure}[t]
\setlength{\abovecaptionskip}{0.cm}
	\begin{center}
        \subfigure
        {\begin{minipage}[b]{.3\linewidth}
        \centering
        \includegraphics[width=2.7cm]{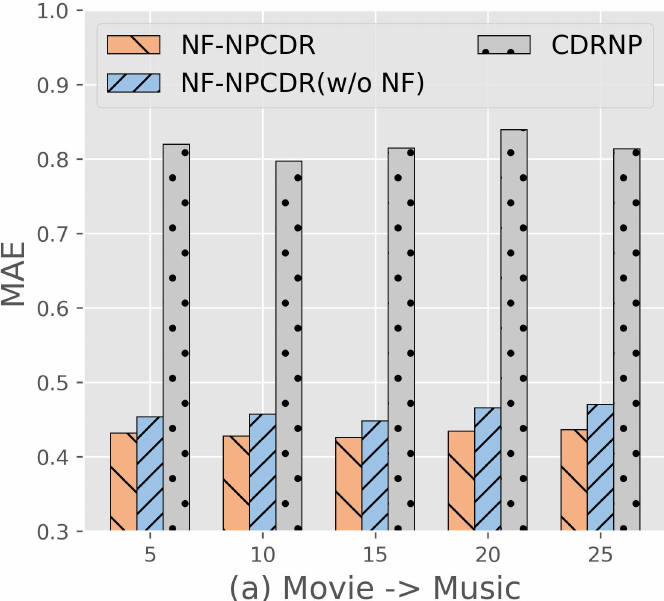}
        \end{minipage}}
        \subfigure
        {\begin{minipage}[b]{.3\linewidth}
        \centering
        \includegraphics[width=2.7cm]{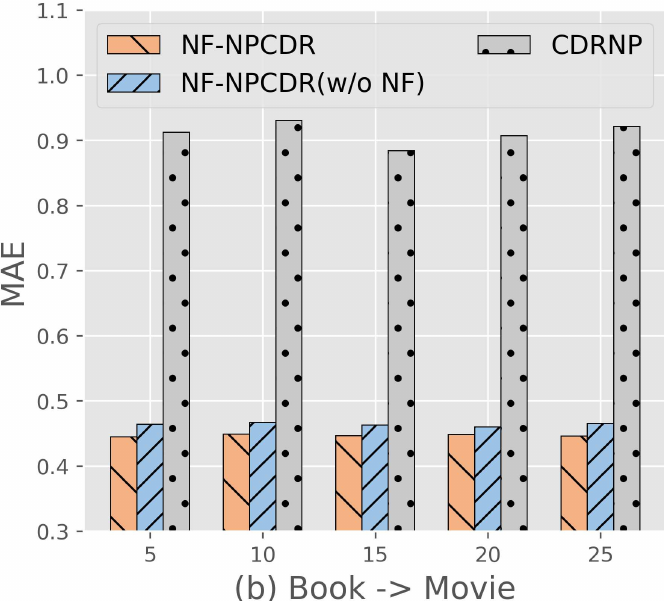}
        \end{minipage}}
        \subfigure
        {\begin{minipage}[b]{.3\linewidth}
        \centering
        \includegraphics[width=2.7cm]{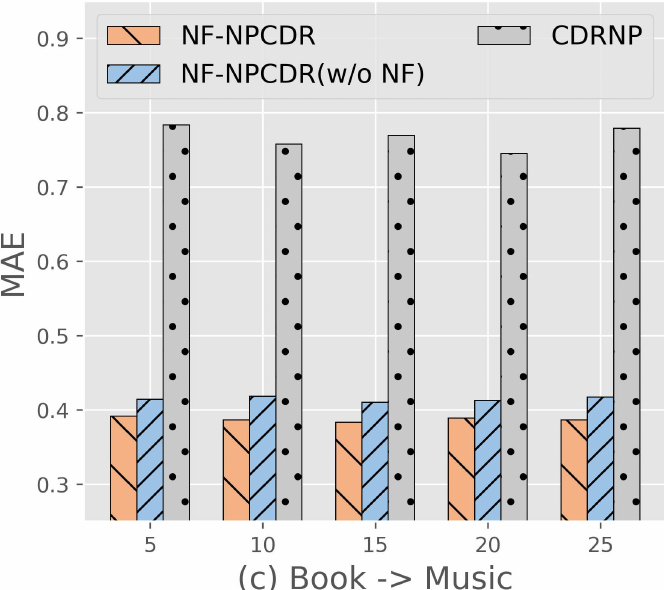}
        \end{minipage}}
        
	\caption{Performance comparison of NF-NPCDR, NF-NPCDR (\textit{w}/\textit{o} NF) and CDRNP with different lengths of support set $\mathcal{C}_i$ on Amazon dataset.}
	\label{entropy_1}
	\end{center}
\vspace{-0.2cm}
\end{figure}

\begin{figure}[t]
\setlength{\abovecaptionskip}{0.cm}
	\begin{center}
        \subfigure
        {\begin{minipage}[b]{.3\linewidth}
        \centering
        \includegraphics[width=2.7cm]{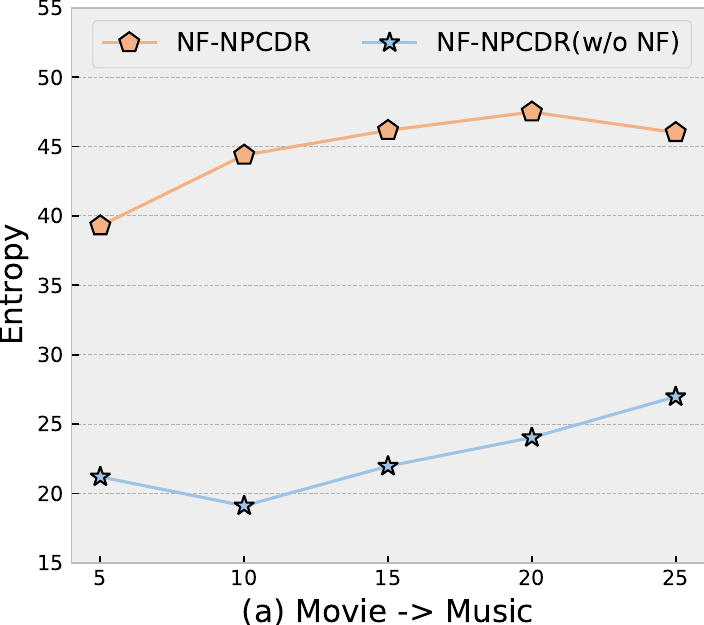}
        \end{minipage}}
        \subfigure
        {\begin{minipage}[b]{.3\linewidth}
        \centering
        \includegraphics[width=2.7cm]{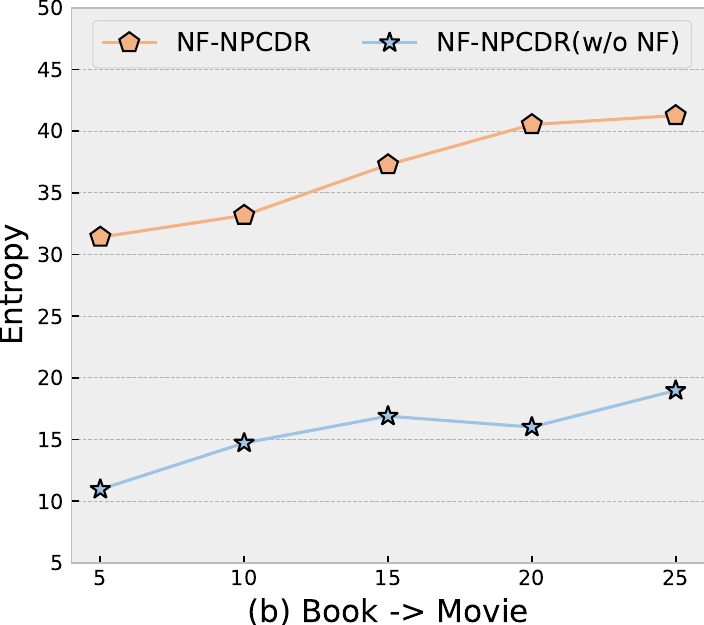}
        \end{minipage}}
        \subfigure
        {\begin{minipage}[b]{.3\linewidth}
        \centering
        \includegraphics[width=2.7cm]{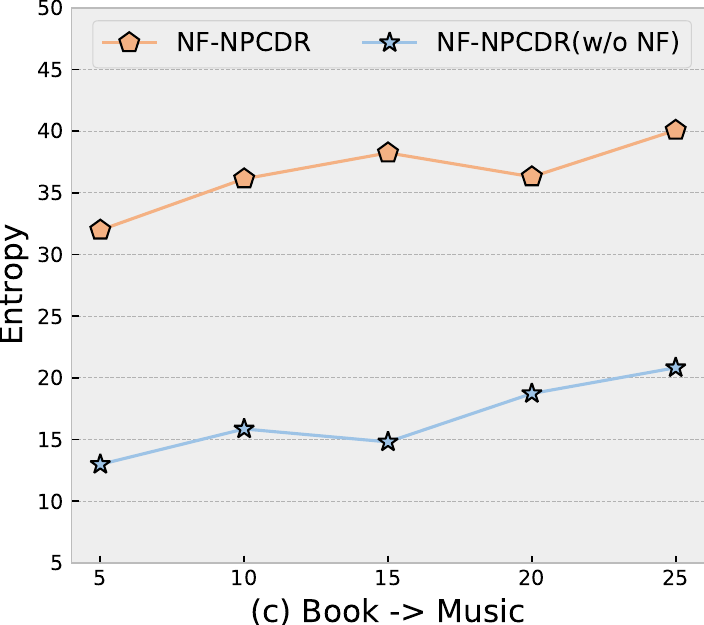}
        \end{minipage}}
        
	\caption{Entropy($\bm{z}_i$) estimated by NF-NPCDR and NF-NPCDR (\textit{w}/\textit{o} NF) with different lengths of support set $\mathcal{C}_i$ on Amazon dataset.}
	\label{entropy_2}
	\end{center}
\vspace{-0.5cm}
\end{figure}

\subsection{Multimodal Distribution Analysis (RQ2)}\label{Multimodal Distribution}

In this section, we further investigate the ability of NF-NPCDR to model a multimodal distribution of CDR prediction functions, which captures the user's personalized multi-interest preference. The distribution of CDR prediction functions can be reflected by the entropy of $\bm{z}_i$. As found by Luo et al.~\cite{npfkgc}, the multimodal distribution leads to a higher entropy than the Gaussian (unimodal) distribution. Therefore, the entropy has a positive correlation with the characteristics of the generated multimodal distribution.
The higher the entropy, the richer the multimodal distribution. Specifically, we first evaluate the performance of NF-NPCDR and NF-NPCDR ($\textit{w}$/$\textit{o}$ NF) with different lengths of support set $\mathcal{C}_i$ in Fig.~\ref{entropy_1}, then illustrate the corresponding entropy estimated by them in Fig.~\ref{entropy_2}. 

From the results shown in Fig~\ref{entropy_1}, we can see that NF-NPCDR consistently outperforms NF-NPCDR ($\textit{w}$/$\textit{o}$ NF) with different lengths of support set $\mathcal{C}_i$. The reason is that normalizing flow can effectively model a multimodal distribution to capture the user's multi-interest preference compared to neural process. This phenomenon can also be illustrated in Fig.~\ref{entropy_2}, the Entropy($\bm{z}_{i,K}$) of NF-NPCDR is always higher than the Entropy($\bm{z}_{i,0}$) of NF-NPCDR ($\textit{w}$/$\textit{o}$ NF). This means that the multimodal distribution can capture richer and more meaningful preference information. Another observation is that both NF-NPCDR and NF-NPCDR ($\textit{w}$/$\textit{o}$ NF) maintain stable performance when the interactions in $\mathcal{C}_i$ is decreased. This indicates that our neural process-based meta-learning paradigm is less sensitive to the decrease of interactions in $\mathcal{C}_i$. This robustness is likely attributable to the use of neural process paradigm, which applies stochastic process to model a Gaussian distribution over user's preference, making our neural process-based meta-learning paradigm more stable.


\begin{table*}[h]
\centering
\scriptsize
\caption{A case study of a cold-start user for CDR from Amazon-Book to Amazon-Movie.}
\setlength{\tabcolsep}{2pt}
\label{case_study1}
\begin{tabular*}{0.85 \textwidth}{@{\extracolsep{\fill}}@{}c|cccccc@{}}
\toprule
\multirow{2}{*}{\diagbox{Amazon-Movie}{Amazon-Book}} & \multicolumn{2}{c}{\color{orange} Interest 1: Adventure, Number: 4} & \multicolumn{2}{c}{\color{blue} Interest 2: Romantic, Number: 3}  & \multicolumn{2}{c}{\color{red} Interest 3: Fantasy, Number: 10}    \\
& \multicolumn{2}{c}{\color{orange} E.g., The Mystery of the Burnt Cottage} & \multicolumn{2}{c}{\color{blue} E.g., Clouds Among the Stars}  & \multicolumn{2}{c}{\color{red} E.g., Dragon Mage}   \\
\midrule
NF-NPCDR (Ours) & ~~~~~\multirow{3}{*}{\includegraphics[width=0.65cm]{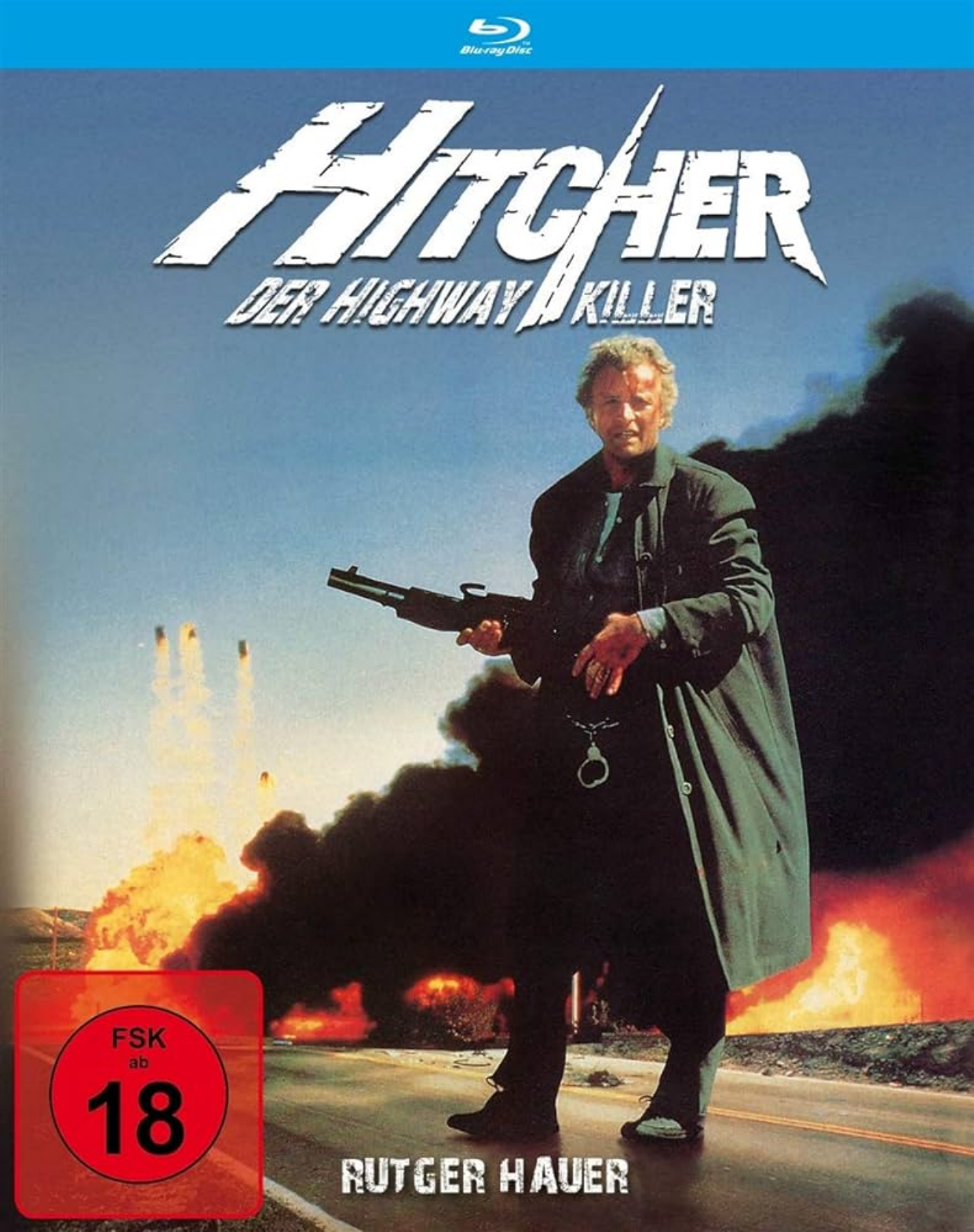}} &  \multirow{4}{*}{\includegraphics[width=1.3cm]{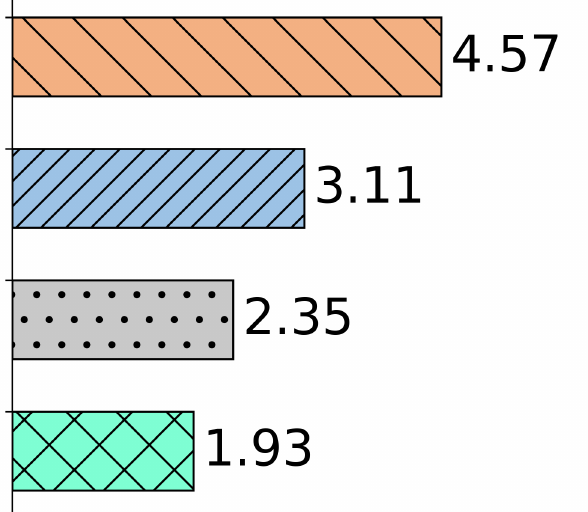}}  &  \multirow{3}{*}{\includegraphics[width=0.55cm]{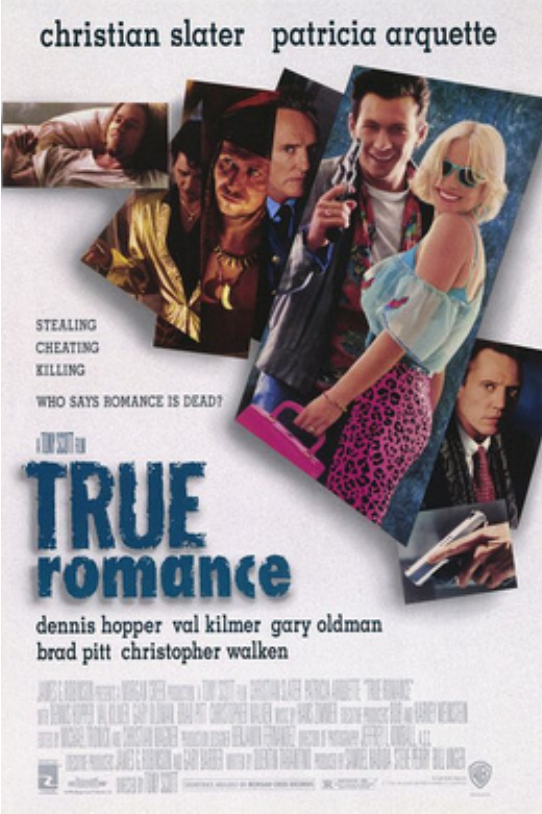}}  &  \multirow{4}{*}{\includegraphics[width=1.3cm]{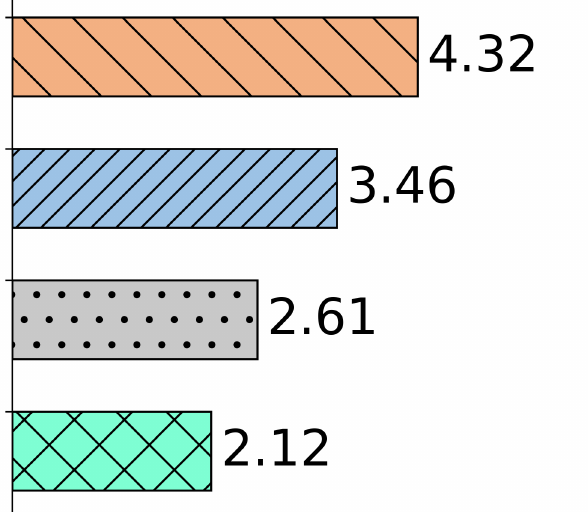}} & \multirow{3}{*}{\includegraphics[width=0.7cm]{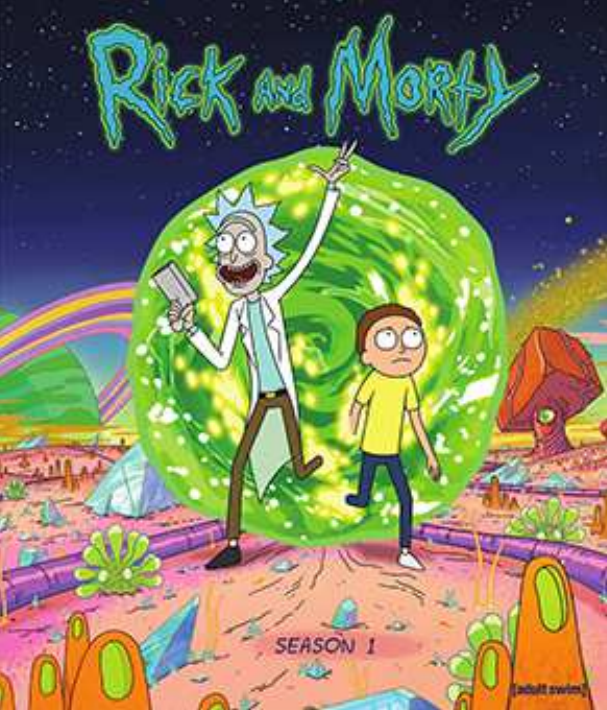}}  &  \multirow{4}{*}{\includegraphics[width=1.3cm]{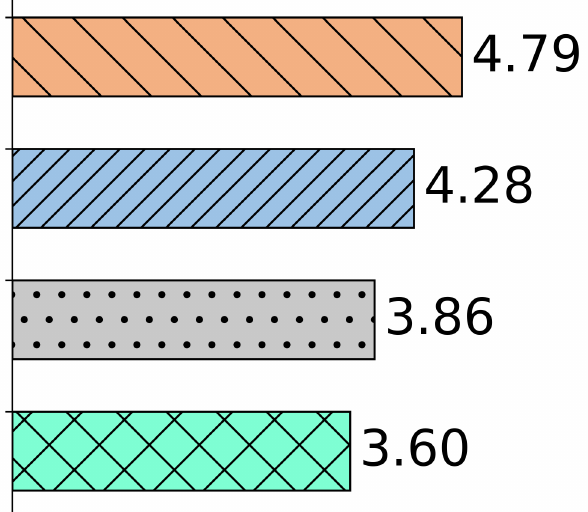}}   \\
NF-NPCDR w/o NF &  &   &   &    &  &     \\
CDRNP &  &    &  &    &  &      \\
PTUPCDR & ~~~~~\color{orange} The Hitcher &   & \color{blue} True Romance &    & \color{red} Rick and Morty &       \\

\bottomrule
\end{tabular*}
\vspace{-0.5cm}
\end{table*}

\subsection{Case Study for Multi-Interest Preference (RQ2)}\label{Case Study}
To better understand the user's personalized multi-interest preference captured by the normalizing flow, we conduct a case study. 
Specifically, we randomly select a cold-start user from the Amazon-Book (source domain) who has no interactions in the Amazon-Movie (target domain). The three interests preference of this cold-start user from the Book domain are Adventure (four books), Romantic (three books) and Fantasy (ten books), represented by orange, blue and red, respectively. And the ratings of this cold-start user to these books are all 5, which are given in the datasets. Finally, the ratings of three candidate movies belonging to Adventure, Romantic and Fantasy are predicted by models.


From the results shown in Table~\ref{case_study1}, we find that our NF-NPCDR achieves the highest ratings in all three candidate movies (e.g., 4.57 on Adventure, 4.32 on Romantic, 4.79 on Fantasy).
On the contrary, NF-NPCDR (\textit{w}/\textit{o} NF) only achieves high score on the Fantasy candidate movie (e.g., 3.11 on Adventure, 3.46 on Romantic, 4.28 on Fantasy). We also note that there are similar observations on CDRNP and PTUPCDR.
The above observations indicate that our NF-NPCDR can capture user's personalized multi-interest preference (including Adventure, Romantic and Fantasy), even though the user interacts with fewer Adventure and Romantic books in the source domain.
This phenomenon can also be explained by the fact that without the normalizing flow, NF-NPCDR (\textit{w}/\textit{o} NF) only captures the user's Fantasy interest that the user interacts with most in the source domain.


\begin{figure}[t]
\setlength{\abovecaptionskip}{0.cm}
	\begin{center}
        \subfigure
        {\begin{minipage}[b]{.3\linewidth}
        \centering
        \includegraphics[width=2.7cm]{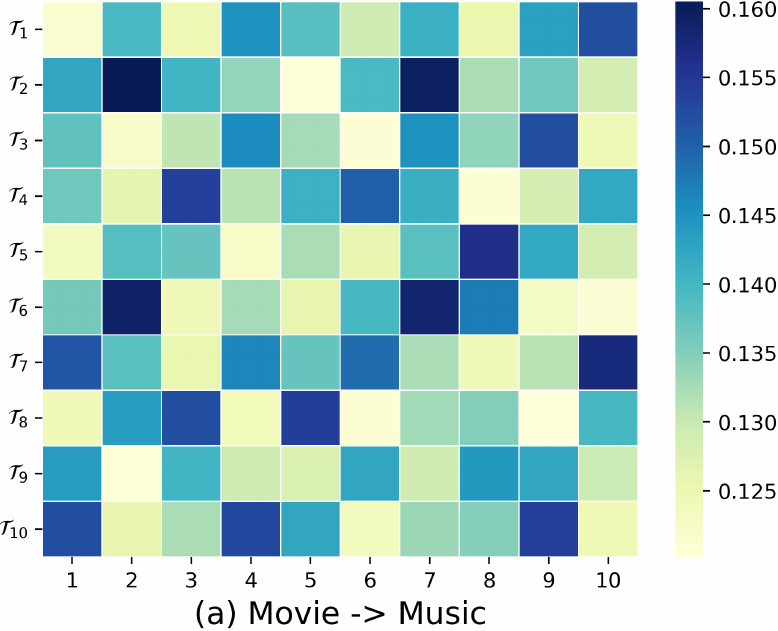}
        \end{minipage}}
        \subfigure
        {\begin{minipage}[b]{.3\linewidth}
        \centering
        \includegraphics[width=2.7cm]{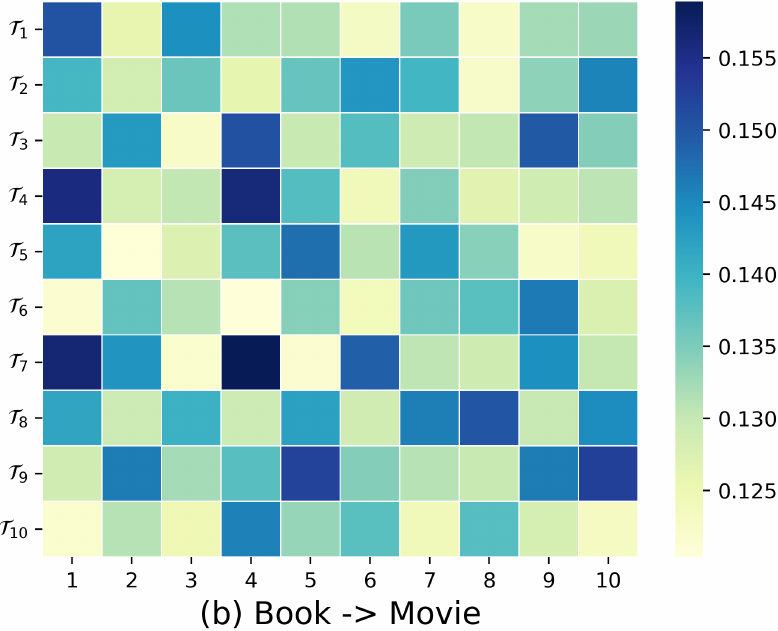}
        \end{minipage}}
        \subfigure
        {\begin{minipage}[b]{.3\linewidth}
        \centering
        \includegraphics[width=2.7cm]{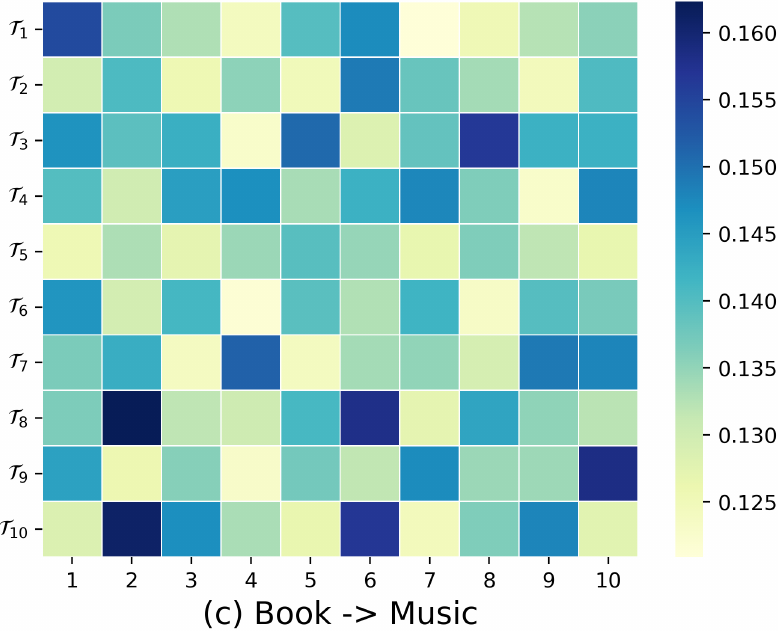}
        \end{minipage}}
        
	\caption{Visualization of soft cluster assignments of 10 users on Amazon dataset. The horizontal axis denotes $N = 10$ soft cluster centroids. If two users achieve high scores (i.e., dark color) on the same soft cluster centroids, they tend to have common preference.}
	\label{visualization}
	\end{center}
\vspace{-0.7cm}
\end{figure}

\subsection{Visualization of Preference Pool (RQ3)}\label{Visualization of preference pool}

In this section, we further visualize the results of soft cluster assignments derived from the interaction between users in $\mathcal{T}_i$ and the preference pool $\mathcal{P}$. Specifically, we randomly sample ten users from the training tasks $\Omega^{tr}$, each user would interact with the preference pool to generate the appropriate soft cluster assignments. If two users achieve high scores (i.e., dark color) on the same soft cluster centroids, they tend to have common preference. On the contrary, if the soft cluster assignments of two users are different, they tend to have dissimilar preference.

The visualization results are shown in Fig.~\ref{visualization}. we observe that the preference pool proposed in our model can capture the common preference between different users. For example, the second and the seventh clusters of $\mathcal{T}_2$ and $\mathcal{T}_6$ in Movie $\to$ Music are simultaneously assigned the highest probability,
which indicates that these two users have similar preference. Meanwhile, the side information also provides proof. We find that these two users both interact with The Lord of the Rings in the source domain. We also have similar findings in other two CDR scenarios. Specifically, we observe that the first and fourth clusters of $\mathcal{T}_4$ and $\mathcal{T}_7$ in Book $\to$ Movie are simultaneously assigned the highest probability,
as well as the second and sixth clusters of $\mathcal{T}_8$ and $\mathcal{T}_{10}$ in Book $\to$ Music,
which means that the preference pool can model common preference between different users. Another observation is that users with dissimilar preference are also distinguish well. For example, $\mathcal{T}_9$ and $\mathcal{T}_{10}$ in Book $\to$ Music have quite different soft cluster assignments, which indicates that these two users have dissimilar preference. From the side information, we find that the interactions of these two users in the source domain do not overlap. Similarly, the soft cluster assignments of $\mathcal{T}_5$ and $\mathcal{T}_{10}$ in Book $\to$ Movie, as well as $\mathcal{T}_5$ and $\mathcal{T}_{7}$ in Book $\to$ Music are quite different. This demonstrates that these users have dissimilar preference.



\subsection{Study of Neural Process (RQ4)}
In this section, we further investigate the impact of neural process in NF-NPCDR ($\textit{w}$/$\textit{o}$ NF) and CDRNP~\cite{cdrnp}. As shown in Fig.~\ref{entropy_1}, we evaluate the performance of NF-NPCDR ($\textit{w}$/$\textit{o}$ NF) and CDRNP with different lengths of support set $\mathcal{C}_i$. We also conduct a case study to investigate the performance of NF-NPCDR ($\textit{w}$/$\textit{o}$ NF) and CDRNP in Fig.~\ref{case_study1}.

From the results shown in Fig.~\ref{entropy_1}, we observe that NF-NPCDR ($\textit{w}$/$\textit{o}$ NF) consistently outperforms CDRNP with different lengths of support set $\mathcal{C}_i$. This demonstrates that directly transferring the users' personalized preference from source to target domain via neural process can achieve better experimental performance. Another observation is that NF-NPCDR ($\textit{w}$/$\textit{o}$ NF) maintains robust performance with different lengths of support set $\mathcal{C}_i$. In contrast, CDRNP is more sensitive to the quality of $\mathcal{C}_i$. The reasons may be that CDRNP splits the interactions of multiple users into one task. There may exist some users with completely different preference, making the model susceptible to noise information.

From the results shown in Table.~\ref{case_study1}, we can also see that the ratings on all three candidate movies of NF-NPCDR ($\textit{w}$/$\textit{o}$ NF) are higher than CDRNP. This indicates that directly leverages the neural process to bridge the source and target domains can achieve significant experimental performance.

\begin{table*}[t]
\centering
\scriptsize
\caption{Analysis on the GPU usage, parameters and training time for one epoch of PTUPCDR, CDRNP and NF-NPCDR on Amazon dataset.}
\setlength{\tabcolsep}{2pt}
\label{computational_analysis}
\begin{tabular*}{0.8 \textwidth}{@{\extracolsep{\fill}}@{}c|ccc@{}}
\toprule
PTUPCDR / CDRNP / NF-NPCDR & Movie $\to$ Music &  Book $\to$ Movie  &  Book $\to$ Music \\
\midrule
\#GPU usage & 1.52G / 1.41G / 1.33G &  2.59G / 2.20G / 1.83G & 2.46G / 2.10G / 1.80G \\
\#Parameters & 25.33M / 18.02M / 11.02M &  84.05M / 66.77M / 38.39M &  80.19M / 65.95M / 37.98M  \\
\#Training time & 307.6s / 56.2s / 177.1s &  583.1s / 98.5s / 364.6s &  288.9s / 43.3s / 168.0s \\

\bottomrule
\end{tabular*}
\vspace{-0.3cm}
\end{table*}

\begin{figure}[t]
\setlength{\abovecaptionskip}{0.cm}
	\begin{center}
        \subfigure
        {\begin{minipage}[b]{.3\linewidth}
        \centering
        \includegraphics[width=2.7cm]{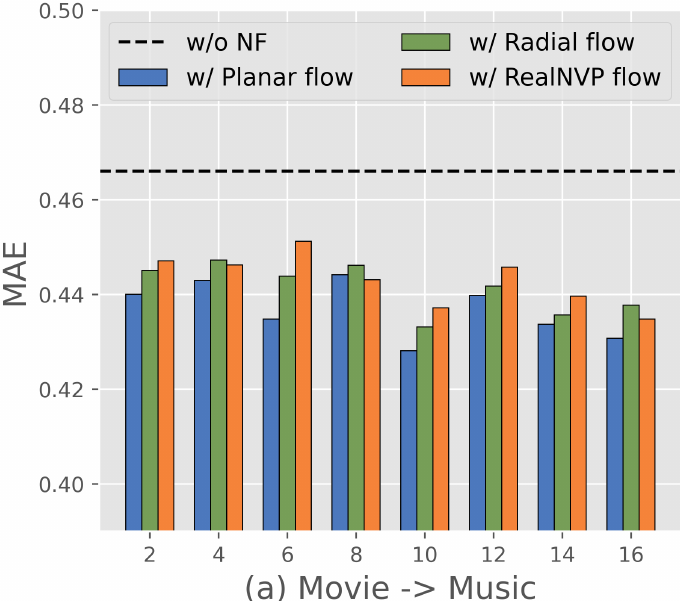}
        \end{minipage}}
        \subfigure
        {\begin{minipage}[b]{.3\linewidth}
        \centering
        \includegraphics[width=2.7cm]{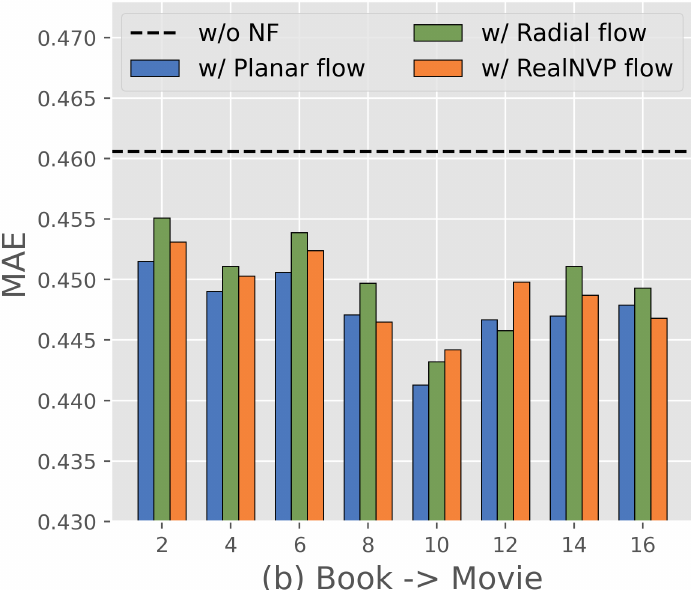}
        \end{minipage}}
        \subfigure
        {\begin{minipage}[b]{.3\linewidth}
        \centering
        \includegraphics[width=2.7cm]{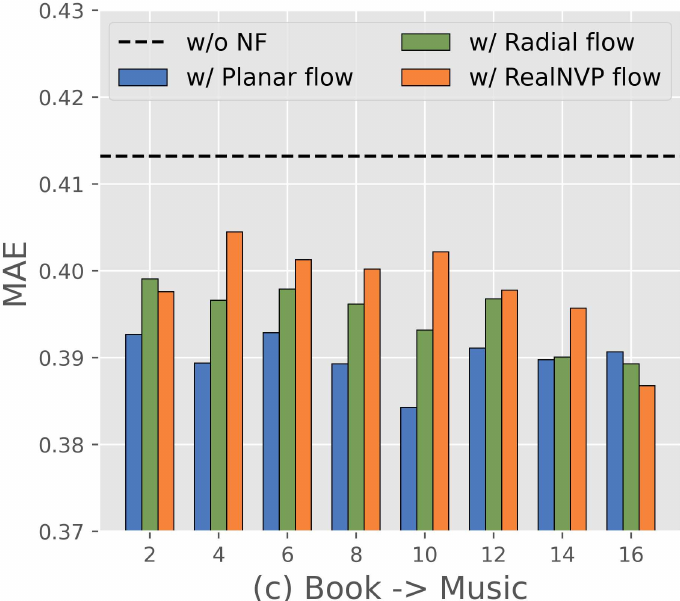}
        \end{minipage}}
        \subfigure
        {\begin{minipage}[b]{.3\linewidth}
        \centering
        \includegraphics[width=2.7cm]{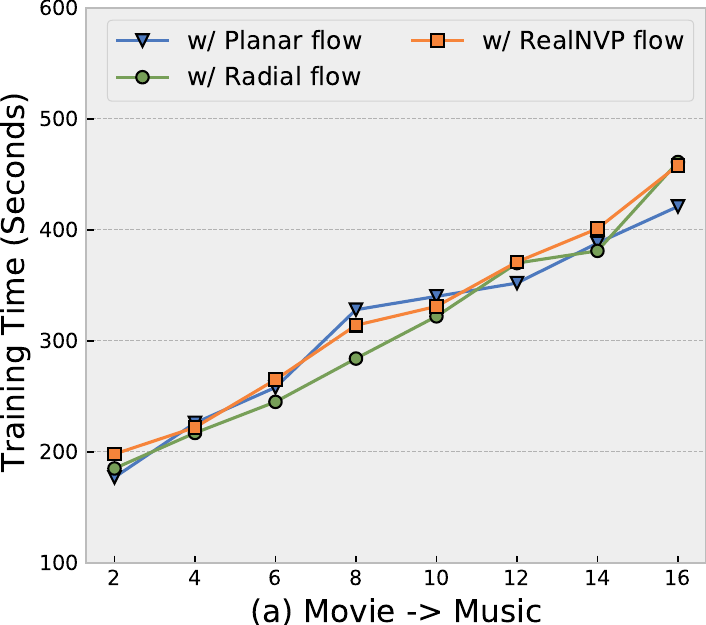}
        \end{minipage}}
        \subfigure
        {\begin{minipage}[b]{.3\linewidth}
        \centering
        \includegraphics[width=2.7cm]{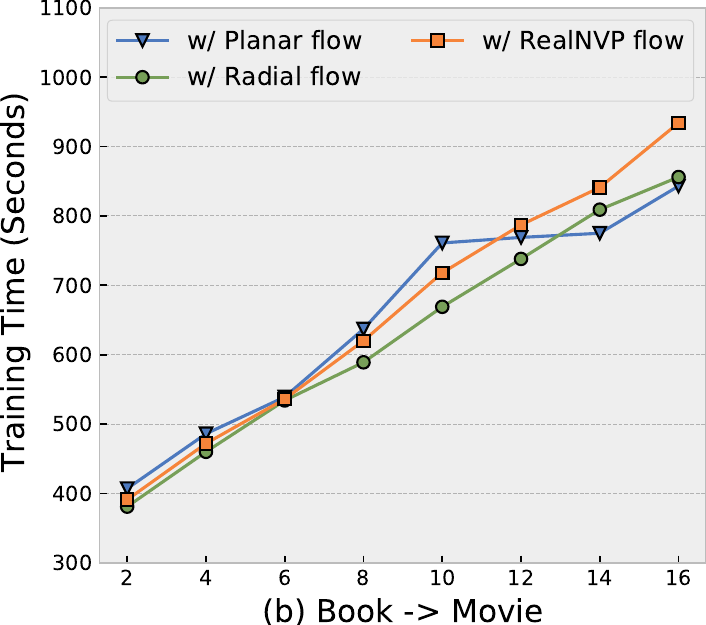}
        \end{minipage}}
        \subfigure
        {\begin{minipage}[b]{.3\linewidth}
        \centering
        \includegraphics[width=2.7cm]{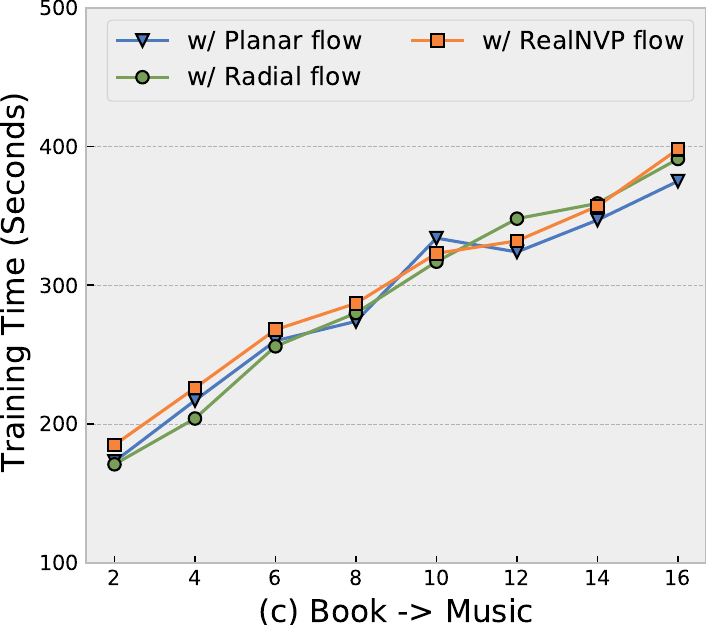}
        \end{minipage}}
        
	\caption{Performance and training time (seconds) for one epoch of NF-NPCDR under different flow steps $K$ on Amazon dataset.}
	\label{running_time}
	\end{center}
\vspace{-0.5cm}
\end{figure}

\subsection{Study of Normalizing Flow (RQ5)}\label{experiment_nf}

We conduct a comprehensive comparative studies on the performance and computational costs of the three normalizing flows (\textit{i.e.}, Planar, Radial and RealNVP). 
Specifically, we explore the NF-NPCDR performance and the training time (seconds) for one epoch with different flow steps $K$.
From the results shown in Fig.~\ref{running_time},
we find that 
compared with \textit{w}/\textit{o} NF variant, 
all three normalizing flows could bring performance improvements, which demonstrates the importance of normalizing flow to capture the user's personalized multi-interest preference.
Another observation is that
NF-NPCDR \textit{w}/ Planar flow consistently outperforms the other flows in all three CDR scenarios with different flow steps, while their computational costs (\textit{i.e.}, training time) are close. This observation indicates that a brief normalizing flow is sufficient to effectively capture user's multi-interest preference in CDR. Therefore, we choose Planar flow in our experiments, which is effective and efficient.
We further introspect the results and observe that the performance of NF-NPCDR (all three normalizing flow variants) improves as the number of steps $K$ increases. This observation indicates that the normalizing flows are able to model a multimodal distribution to capture the user's multi-interest preference with increased steps $K$. We also find that while normalizing flows enhance the effectiveness of the neural process, the computational expense also rises with an increase in the number of steps $K$. The results shown in Fig.~\ref{running_time} clearly demonstrate that a gradual increase in training time as the steps $K$ increase. Through the above observations, we set the steps $K$ of normalizing (Planar) flows to six in Movie $\to$ Music but four in Book $\to$ Movie and Book $\to$ Music. The reason is that, incorporating an excessive number of flows not only slows down computation but also risks overfitting. Thus, selecting a computationally brief flow or using a smaller number of steps $K$ to further enhance the computational efficiency.




\subsection{Computational Cost Analysis (RQ6)}

We further provide the computational cost analysis of the time and space costs for our NF-NPCDR and the classical method PTUPCDR~\cite{ptupcdr}, as well as the SOTA method CDRNP~\cite{cdrnp}, including GPU usage, parameters and training time for one epoch. From the results shown in Table~\ref{computational_analysis}, we find that despite the slightly longer training time compared to CDRNP, the efficiency of NF-NPCDR is still in the same order of magnitude as PTUPCDR. Moreover, the GPU usage and the parameters of NF-NPCDR are much lower than those of PTUPCDR and CDRNP, further demonstrating the scalability and feasibility of our personalized multi-interest modeling framework for large-scale deployment. Besides, considering the significant performance improvements of our NF-NPCDR, the computational costs are reasonable and acceptable.

\begin{figure}[t]
\setlength{\abovecaptionskip}{0.cm}
	\begin{center}
        \subfigure
        {\begin{minipage}[b]{.4\linewidth}
        \centering
        \includegraphics[width=2.7cm]{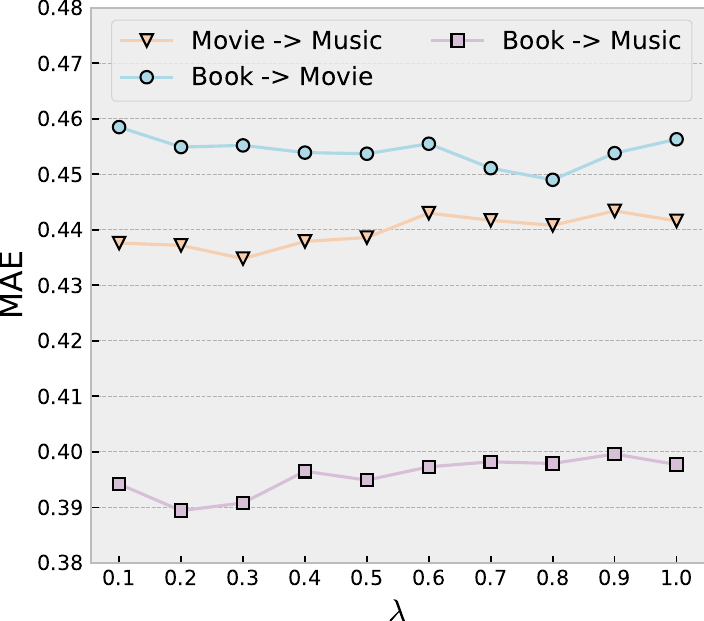}
        \end{minipage}}
        \subfigure
        {\begin{minipage}[b]{.4\linewidth}
        \centering
        \includegraphics[width=2.7cm]{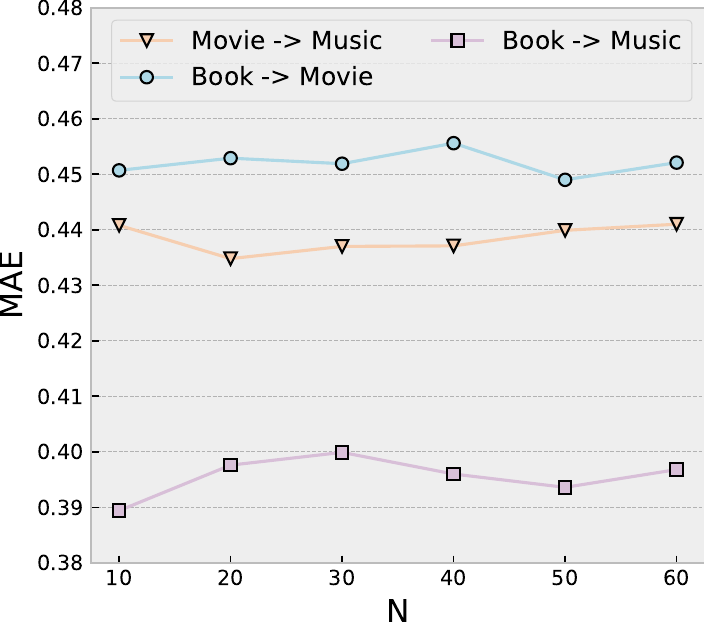}
        \end{minipage}}
        
	\caption{Sensitivity of NF-NPCDR to hyper-parameters $\lambda$ and $N$ on Amazon.}
	\label{hyper-parameter}
	\end{center}
\vspace{-0.5cm}
\end{figure}

        

\subsection{Hyper-parameter Analysis}
We explore the effect of different hyper-parameter settings to our NF-NPCDR in this section, e.g., $\lambda$ in Eq.~(\ref{total_loss}) and the number of soft cluster centroids $N$ in the preference pool $\mathcal{P}$. As the results shown in Fig.~\ref{hyper-parameter}, we observe that in scenario with extensive interactions, such as the Amazon-Book $\to$ Amazon-Movie, $\lambda$ and $N$ can be assigned a larger value. This may be due to the fact that extensive interactions scenario typically involves more complex common preference between different users,
which should be captured by larger $\lambda$ and $N$. In contrast, for the Amazon-Movie $\to$ Amazon-Music and Amazon-Book $\to$ Amazon-Music, smaller $\lambda$ and $N$ can achieve good recommendation performance. We also find that NF-NPCDR maintains robust performance with different values of $\lambda$ and $N$. This can be attributed to our personalized multi-interest modeling framework, which enhances the neural process with the normalizing flow to convert the Gaussian (unimodal) distribution to a multimodal distribution, making our model more robust and stable to the hyper-parameters.

\section{Conclusion}\label{conclusion}

In this paper, we propose a novel personalized multi-interest modeling framework for CDR to cold-start users.
We first enhance the neural process with the normalizing flow to convert the Gaussian (unimodal) distribution to a multimodal distribution, which captures the user's personalized multi-interest preference. Then, we propose a common preference encoder with a preference pool to capture the common preference between different users. Furthermore, we introduce a stochastic adaptive decoder to further incorporate both the personalized and common preference for cold-start users, adaptively modulating both preference for better recommendation. Extensive experiments demonstrate that NF-NPCDR consistently outperforms the previous SOTA approaches in five real-world CDR scenarios. 
\footnotesize
\bibliographystyle{IEEEtranN} 
\bibliography{IEEEabrv,mybibfile}

\begin{thebibliography}{70}
\providecommand{\natexlab}[1]{#1}
\providecommand{\url}[1]{#1}
\csname url@samestyle\endcsname
\providecommand{\newblock}{\relax}
\providecommand{\bibinfo}[2]{#2}
\providecommand{\BIBentrySTDinterwordspacing}{\spaceskip=0pt\relax}
\providecommand{\BIBentryALTinterwordstretchfactor}{4}
\providecommand{\BIBentryALTinterwordspacing}{\spaceskip=\fontdimen2\font plus
\BIBentryALTinterwordstretchfactor\fontdimen3\font minus \fontdimen4\font\relax}
\providecommand{\BIBforeignlanguage}[2]{{%
\expandafter\ifx\csname l@#1\endcsname\relax
\typeout{** WARNING: IEEEtranN.bst: No hyphenation pattern has been}%
\typeout{** loaded for the language `#1'. Using the pattern for}%
\typeout{** the default language instead.}%
\else
\language=\csname l@#1\endcsname
\fi
#2}}
\providecommand{\BIBdecl}{\relax}
\BIBdecl

\bibitem[Rendle et~al.(2012)Rendle, Freudenthaler, Gantner, and Schmidt-Thieme]{BPR}
S.~Rendle, C.~Freudenthaler, Z.~Gantner, and L.~Schmidt-Thieme, ``Bpr: Bayesian personalized ranking from implicit feedback,'' \emph{arXiv preprint arXiv:1205.2618}, 2012.

\bibitem[Singh and Gordon(2008)]{cmf}
A.~P. Singh and G.~J. Gordon, ``Relational learning via collective matrix factorization,'' in \emph{Proceedings of the 14th ACM SIGKDD international conference on Knowledge discovery and data mining}, 2008, pp. 650--658.

\bibitem[He et~al.(2017)He, Liao, Zhang, Nie, Hu, and Chua]{ncf}
X.~He, L.~Liao, H.~Zhang, L.~Nie, X.~Hu, and T.-S. Chua, ``Neural collaborative filtering,'' in \emph{Proceedings of the 26th international conference on world wide web}, 2017, pp. 173--182.

\bibitem[Wang et~al.(2019)Wang, He, Wang, Feng, and Chua]{ngcf}
X.~Wang, X.~He, M.~Wang, F.~Feng, and T.-S. Chua, ``Neural graph collaborative filtering,'' in \emph{Proceedings of the 42nd international ACM SIGIR conference on Research and development in Information Retrieval}, 2019, pp. 165--174.

\bibitem[Lin et~al.(2021)Lin, Wu, Zhou, Pan, Cao, and Wang]{tanp}
X.~Lin, J.~Wu, C.~Zhou, S.~Pan, Y.~Cao, and B.~Wang, ``Task-adaptive neural process for user cold-start recommendation,'' in \emph{Proceedings of the Web Conference 2021}, 2021, pp. 1306--1316.

\bibitem[Kang et~al.(2019)Kang, Hwang, Lee, and Yu]{sscdr}
S.~Kang, J.~Hwang, D.~Lee, and H.~Yu, ``Semi-supervised learning for cross-domain recommendation to cold-start users,'' in \emph{Proceedings of the 28th ACM International Conference on Information and Knowledge Management}, 2019, pp. 1563--1572.

\bibitem[Man et~al.(2017)Man, Shen, Jin, and Cheng]{emcdr}
T.~Man, H.~Shen, X.~Jin, and X.~Cheng, ``Cross-domain recommendation: An embedding and mapping approach.'' in \emph{IJCAI}, vol.~17, 2017, pp. 2464--2470.

\bibitem[Salah et~al.(2021)Salah, Tran, and Lauw]{tsvcdr}
A.~Salah, T.~B. Tran, and H.~Lauw, ``Towards source-aligned variational models for cross-domain recommendation,'' in \emph{Proceedings of the 15th ACM Conference on Recommender Systems}, 2021, pp. 176--186.

\bibitem[Shi and Wang(2019)]{cdrva}
J.~Shi and Q.~Wang, ``Cross-domain variational autoencoder for recommender systems,'' in \emph{2019 IEEE 11th International Conference on Advanced Infocomm Technology (ICAIT)}.\hskip 1em plus 0.5em minus 0.4em\relax IEEE, 2019, pp. 67--72.

\bibitem[Zhu et~al.(2021)Zhu, Ge, Zhuang, Xie, Xi, Zhang, Lin, and He]{tmcdr}
Y.~Zhu, K.~Ge, F.~Zhuang, R.~Xie, D.~Xi, X.~Zhang, L.~Lin, and Q.~He, ``Transfer-meta framework for cross-domain recommendation to cold-start users,'' in \emph{Proceedings of the 44th International ACM SIGIR Conference on Research and Development in Information Retrieval}, 2021, pp. 1813--1817.

\bibitem[Zhu et~al.(2018)Zhu, Wang, Chen, Liu, Orgun, and Wu]{dcdcsr}
F.~Zhu, Y.~Wang, C.~Chen, G.~Liu, M.~Orgun, and J.~Wu, ``A deep framework for cross-domain and cross-system recommendations,'' in \emph{Proceedings of the 27th International Joint Conference on Artificial Intelligence}, 2018, pp. 3711--3717.

\bibitem[Bi et~al.(2020{\natexlab{a}})Bi, Song, Yao, Wu, Wang, and Xiao]{dcdir}
Y.~Bi, L.~Song, M.~Yao, Z.~Wu, J.~Wang, and J.~Xiao, ``Dcdir: A deep cross-domain recommendation system for cold start users in insurance domain,'' in \emph{Proceedings of the 43rd international ACM SIGIR conference on research and development in information retrieval}, 2020, pp. 1661--1664.

\bibitem[Bi et~al.(2020{\natexlab{b}})Bi, Song, Yao, Wu, Wang, and Xiao]{hcdir}
------, ``A heterogeneous information network based cross domain insurance recommendation system for cold start users,'' in \emph{Proceedings of the 43rd international ACM SIGIR conference on research and development in information retrieval}, 2020, pp. 2211--2220.

\bibitem[Zhu et~al.(2022)Zhu, Tang, Liu, Zhuang, Xie, Zhang, Lin, and He]{ptupcdr}
Y.~Zhu, Z.~Tang, Y.~Liu, F.~Zhuang, R.~Xie, X.~Zhang, L.~Lin, and Q.~He, ``Personalized transfer of user preferences for cross-domain recommendation,'' in \emph{Proceedings of the Fifteenth ACM International Conference on Web Search and Data Mining}, 2022, pp. 1507--1515.

\bibitem[Guan et~al.(2022)Guan, Pang, Giunchiglia, Liang, and Feng]{cdml}
R.~Guan, H.~Pang, F.~Giunchiglia, Y.~Liang, and X.~Feng, ``Cross-domain meta-learner for cold-start recommendation,'' \emph{IEEE Transactions on Knowledge and Data Engineering}, 2022.

\bibitem[Cen et~al.(2020{\natexlab{a}})Cen, Zhang, Zou, Zhou, Yang, and Tang]{cmiffr}
Y.~Cen, J.~Zhang, X.~Zou, C.~Zhou, H.~Yang, and J.~Tang, ``Controllable multi-interest framework for recommendation,'' in \emph{Proceedings of the 26th ACM SIGKDD International Conference on Knowledge Discovery \& Data Mining}, 2020, pp. 2942--2951.

\bibitem[Chen et~al.(2021)Chen, Zhang, Zhao, Xue, and Xiang]{epimi}
G.~Chen, X.~Zhang, Y.~Zhao, C.~Xue, and J.~Xiang, ``Exploring periodicity and interactivity in multi-interest framework for sequential recommendation,'' \emph{Proceedings of the 30th International Joint Conference on Artificial Intelligence}, 2021.

\bibitem[Garnelo et~al.(2018{\natexlab{a}})Garnelo, Schwarz, Rosenbaum, Viola, Rezende, Eslami, and Teh]{np}
M.~Garnelo, J.~Schwarz, D.~Rosenbaum, F.~Viola, D.~J. Rezende, S.~Eslami, and Y.~W. Teh, ``Neural processes,'' \emph{arXiv preprint arXiv:1807.01622}, 2018.

\bibitem[Papamakarios et~al.(2021)Papamakarios, Nalisnick, Rezende, Mohamed, and Lakshminarayanan]{nf}
G.~Papamakarios, E.~Nalisnick, D.~J. Rezende, S.~Mohamed, and B.~Lakshminarayanan, ``Normalizing flows for probabilistic modeling and inference,'' \emph{The Journal of Machine Learning Research}, vol.~22, no.~1, pp. 2617--2680, 2021.

\bibitem[Tabak and Turner(2013)]{ndea}
E.~G. Tabak and C.~V. Turner, ``A family of nonparametric density estimation algorithms,'' \emph{Communications on Pure and Applied Mathematics}, vol.~66, no.~2, pp. 145--164, 2013.

\bibitem[Li et~al.(2021)Li, Yao, Mu, Zhao, Li, Guo, Ding, and Wen]{dlcdr}
S.~Li, L.~Yao, S.~Mu, W.~X. Zhao, Y.~Li, T.~Guo, B.~Ding, and J.-R. Wen, ``Debiasing learning based cross-domain recommendation,'' in \emph{Proceedings of the 27th ACM SIGKDD Conference on Knowledge Discovery \& Data Mining}, 2021, pp. 3190--3199.

\bibitem[Fan et~al.(2021)Fan, Derr, Zhao, Ma, Liu, Wang, Tang, and Li]{abrccdr}
W.~Fan, T.~Derr, X.~Zhao, Y.~Ma, H.~Liu, J.~Wang, J.~Tang, and Q.~Li, ``Attacking black-box recommendations via copying cross-domain user profiles,'' in \emph{2021 IEEE 37th International Conference on Data Engineering (ICDE)}.\hskip 1em plus 0.5em minus 0.4em\relax IEEE, 2021, pp. 1583--1594.

\bibitem[Li and Tuzhilin(2021)]{dmlee}
P.~Li and A.~Tuzhilin, ``Dual metric learning for effective and efficient cross-domain recommendations,'' \emph{IEEE Transactions on Knowledge and Data Engineering}, vol.~35, no.~1, pp. 321--334, 2021.

\bibitem[Zhao et~al.(2023)Zhao, Zhao, Li, He, Wang, and Fan]{cdrpsa}
C.~Zhao, H.~Zhao, X.~Li, M.~He, J.~Wang, and J.~Fan, ``Cross-domain recommendation via progressive structural alignment,'' \emph{IEEE Transactions on Knowledge and Data Engineering}, 2023.

\bibitem[Hu et~al.(2018)Hu, Zhang, and Yang]{conet}
G.~Hu, Y.~Zhang, and Q.~Yang, ``Conet: Collaborative cross networks for cross-domain recommendation,'' in \emph{Proceedings of the 27th ACM international conference on information and knowledge management}, 2018, pp. 667--676.

\bibitem[Li et~al.(2009)Li, Yang, and Xue]{cdcfsr}
B.~Li, Q.~Yang, and X.~Xue, ``Can movies and books collaborate? cross-domain collaborative filtering for sparsity reduction,'' in \emph{Twenty-First international joint conference on artificial intelligence}, 2009.

\bibitem[Cao et~al.(2022{\natexlab{a}})Cao, Lin, Cong, Ya, Liu, and Wang]{disencdr}
J.~Cao, X.~Lin, X.~Cong, J.~Ya, T.~Liu, and B.~Wang, ``Disencdr: Learning disentangled representations for cross-domain recommendation,'' in \emph{Proceedings of the 45th International ACM SIGIR Conference on Research and Development in Information Retrieval}, 2022, pp. 267--277.

\bibitem[Liu et~al.(2023)Liu, Zheng, Su, Zheng, Chen, and Hu]{cpkt}
W.~Liu, X.~Zheng, J.~Su, L.~Zheng, C.~Chen, and M.~Hu, ``Contrastive proxy kernel stein path alignment for cross-domain cold-start recommendation,'' \emph{IEEE Transactions on Knowledge and Data Engineering}, 2023.

\bibitem[Cao et~al.(2022{\natexlab{b}})Cao, Sheng, Cong, Liu, and Wang]{cdrib}
J.~Cao, J.~Sheng, X.~Cong, T.~Liu, and B.~Wang, ``Cross-domain recommendation to cold-start users via variational information bottleneck,'' in \emph{2022 IEEE 38th International Conference on Data Engineering (ICDE)}.\hskip 1em plus 0.5em minus 0.4em\relax IEEE, 2022, pp. 2209--2223.

\bibitem[Sun et~al.(2023)Sun, Gu, Hu, Dong, Li, Cheng, and Mo]{remit}
C.~Sun, J.~Gu, B.~Hu, X.~Dong, H.~Li, L.~Cheng, and L.~Mo, ``Remit: reinforced multi-interest transfer for cross-domain recommendation,'' in \emph{Proceedings of the AAAI Conference on Artificial Intelligence}, vol.~37, no.~8, 2023, pp. 9900--9908.

\bibitem[Garnelo et~al.(2018{\natexlab{b}})Garnelo, Rosenbaum, Maddison, Ramalho, Saxton, Shanahan, Teh, Rezende, and Eslami]{cnp}
M.~Garnelo, D.~Rosenbaum, C.~Maddison, T.~Ramalho, D.~Saxton, M.~Shanahan, Y.~W. Teh, D.~Rezende, and S.~A. Eslami, ``Conditional neural processes,'' in \emph{International conference on machine learning}.\hskip 1em plus 0.5em minus 0.4em\relax PMLR, 2018, pp. 1704--1713.

\bibitem[Kim et~al.(2019)Kim, Mnih, Schwarz, Garnelo, Eslami, Rosenbaum, Vinyals, and Teh]{anp}
H.~Kim, A.~Mnih, J.~Schwarz, M.~Garnelo, A.~Eslami, D.~Rosenbaum, O.~Vinyals, and Y.~W. Teh, ``Attentive neural processes,'' \emph{arXiv preprint arXiv:1901.05761}, 2019.

\bibitem[Shen et~al.(2023)Shen, Zhen, Wang, and Worring]{mtlnp}
J.~Shen, X.~Zhen, Q.~Wang, and M.~Worring, ``Episodic multi-task learning with heterogeneous neural processes,'' \emph{Advances in Neural Information Processing Systems}, vol.~36, pp. 75\,214--75\,228, 2023.

\bibitem[Wang et~al.(2023)Wang, Massiceti, Hu, Pavlovic, and Lukasiewicz]{NP-SemiSeg}
J.~Wang, D.~Massiceti, X.~Hu, V.~Pavlovic, and T.~Lukasiewicz, ``Np-semiseg: when neural processes meet semi-supervised semantic segmentation,'' in \emph{International Conference on Machine Learning}.\hskip 1em plus 0.5em minus 0.4em\relax PMLR, 2023, pp. 36\,138--36\,156.

\bibitem[Du et~al.(2023)Du, Ye, Guo, Yu, and Yao]{idnp}
J.~Du, Z.~Ye, B.~Guo, Z.~Yu, and L.~Yao, ``Idnp: Interest dynamics modeling using generative neural processes for sequential recommendation,'' in \emph{Proceedings of the Sixteenth ACM International Conference on Web Search and Data Mining}, 2023, pp. 481--489.

\bibitem[Lin et~al.(2023)Lin, Zhou, Wu, Zou, Pan, Cao, Wang, Wang, and Yin]{tfanp}
X.~Lin, C.~Zhou, J.~Wu, L.~Zou, S.~Pan, Y.~Cao, B.~Wang, S.~Wang, and D.~Yin, ``Towards flexible and adaptive neural process for cold-start recommendation,'' \emph{IEEE Transactions on Knowledge and Data Engineering}, 2023.

\bibitem[Liu et~al.(2022)Liu, Jing, Yu, Zhou, and Ng]{lieinp}
H.~Liu, L.~Jing, D.~Yu, M.~Zhou, and M.~Ng, ``Learning intrinsic and extrinsic intentions for cold-start recommendation with neural stochastic processes,'' in \emph{Proceedings of the 30th ACM International Conference on Multimedia}, 2022, pp. 491--500.

\bibitem[Li et~al.(2024)Li, Sheng, Cao, Zhang, Li, and Liu]{cdrnp}
X.~Li, J.~Sheng, J.~Cao, W.~Zhang, Q.~Li, and T.~Liu, ``Cdrnp: Cross-domain recommendation to cold-start users via neural process,'' in \emph{Proceedings of the 17th ACM International Conference on Web Search and Data Mining}, 2024, pp. 378--386.

\bibitem[Kobyzev et~al.(2020)Kobyzev, Prince, and Brubaker]{nfs}
I.~Kobyzev, S.~J. Prince, and M.~A. Brubaker, ``Normalizing flows: An introduction and review of current methods,'' \emph{IEEE transactions on pattern analysis and machine intelligence}, vol.~43, no.~11, pp. 3964--3979, 2020.

\bibitem[Rezende and Mohamed(2015{\natexlab{a}})]{planarflow}
D.~Rezende and S.~Mohamed, ``Variational inference with normalizing flows,'' in \emph{International conference on machine learning}.\hskip 1em plus 0.5em minus 0.4em\relax PMLR, 2015, pp. 1530--1538.

\bibitem[Dinh et~al.(2016)Dinh, Sohl-Dickstein, and Bengio]{realnvpflow}
L.~Dinh, J.~Sohl-Dickstein, and S.~Bengio, ``Density estimation using real nvp,'' \emph{arXiv preprint arXiv:1605.08803}, 2016.

\bibitem[Papamakarios et~al.(2017)Papamakarios, Pavlakou, and Murray]{mafflow}
G.~Papamakarios, T.~Pavlakou, and I.~Murray, ``Masked autoregressive flow for density estimation,'' \emph{Advances in neural information processing systems}, vol.~30, 2017.

\bibitem[Mazoure et~al.(2020)Mazoure, Doan, Durand, Pineau, and Hjelm]{lenf}
B.~Mazoure, T.~Doan, A.~Durand, J.~Pineau, and R.~D. Hjelm, ``Leveraging exploration in off-policy algorithms via normalizing flows,'' in \emph{Conference on Robot Learning}.\hskip 1em plus 0.5em minus 0.4em\relax PMLR, 2020, pp. 430--444.

\bibitem[Touati et~al.(2020)Touati, Satija, Romoff, Pineau, and Vincent]{rvfnf}
A.~Touati, H.~Satija, J.~Romoff, J.~Pineau, and P.~Vincent, ``Randomized value functions via multiplicative normalizing flows,'' in \emph{Uncertainty in Artificial Intelligence}.\hskip 1em plus 0.5em minus 0.4em\relax PMLR, 2020, pp. 422--432.

\bibitem[Madhawa et~al.(2019)Madhawa, Ishiguro, Nakago, and Abe]{graphnvp}
K.~Madhawa, K.~Ishiguro, K.~Nakago, and M.~Abe, ``Graphnvp: An invertible flow model for generating molecular graphs,'' \emph{arXiv preprint arXiv:1905.11600}, 2019.

\bibitem[Luo et~al.(2023)Luo, Li, Haffari, and Pan]{npfkgc}
L.~Luo, Y.-F. Li, G.~Haffari, and S.~Pan, ``Normalizing flow-based neural process for few-shot knowledge graph completion,'' \emph{arXiv preprint arXiv:2304.08183}, 2023.

\bibitem[Kingma and Welling(2013)]{aevb}
D.~P. Kingma and M.~Welling, ``Auto-encoding variational bayes,'' \emph{arXiv preprint arXiv:1312.6114}, 2013.

\bibitem[Mnih and Gregor(2014)]{nvi}
A.~Mnih and K.~Gregor, ``Neural variational inference and learning in belief networks,'' in \emph{International Conference on Machine Learning}.\hskip 1em plus 0.5em minus 0.4em\relax PMLR, 2014, pp. 1791--1799.

\bibitem[Rezende and Mohamed(2015{\natexlab{b}})]{vinf}
D.~Rezende and S.~Mohamed, ``Variational inference with normalizing flows,'' in \emph{International conference on machine learning}.\hskip 1em plus 0.5em minus 0.4em\relax PMLR, 2015, pp. 1530--1538.

\bibitem[{\O}ksendal and {\O}ksendal(2003)]{stochastic03}
B.~{\O}ksendal and B.~{\O}ksendal, \emph{Stochastic differential equations}.\hskip 1em plus 0.5em minus 0.4em\relax Springer, 2003.

\bibitem[Xie et~al.(2016)Xie, Girshick, and Farhadi]{udeca}
J.~Xie, R.~Girshick, and A.~Farhadi, ``Unsupervised deep embedding for clustering analysis,'' in \emph{International conference on machine learning}.\hskip 1em plus 0.5em minus 0.4em\relax PMLR, 2016, pp. 478--487.

\bibitem[Van~der Maaten and Hinton(2008)]{t_sne}
L.~Van~der Maaten and G.~Hinton, ``Visualizing data using t-sne.'' \emph{Journal of machine learning research}, vol.~9, no.~11, 2008.

\bibitem[Perez et~al.(2018)Perez, Strub, De~Vries, Dumoulin, and Courville]{film}
E.~Perez, F.~Strub, H.~De~Vries, V.~Dumoulin, and A.~Courville, ``Film: Visual reasoning with a general conditioning layer,'' in \emph{Proceedings of the AAAI Conference on Artificial Intelligence}, vol.~32, no.~1, 2018.

\bibitem[Fan et~al.(2022)Fan, Lian, Zhao, Liu, Li, and Xie]{ada}
X.~Fan, J.~Lian, W.~X. Zhao, Z.~Liu, C.~Li, and X.~Xie, ``Ada-ranker: A data distribution adaptive ranking paradigm for sequential recommendation,'' in \emph{Proceedings of the 45th International ACM SIGIR Conference on Research and Development in Information Retrieval}, 2022, pp. 1599--1610.

\bibitem[Cen et~al.(2020{\natexlab{b}})Cen, Zhang, Zou, Zhou, Yang, and Tang]{cmif}
Y.~Cen, J.~Zhang, X.~Zou, C.~Zhou, H.~Yang, and J.~Tang, ``Controllable multi-interest framework for recommendation,'' in \emph{Proceedings of the 26th ACM SIGKDD International Conference on Knowledge Discovery \& Data Mining}, 2020, pp. 2942--2951.

\bibitem[Guo et~al.(2022)Guo, Zhang, He, Qin, Guo, Chen, Tang, He, and Zhang]{miss}
W.~Guo, C.~Zhang, Z.~He, J.~Qin, H.~Guo, B.~Chen, R.~Tang, X.~He, and R.~Zhang, ``Miss: Multi-interest self-supervised learning framework for click-through rate prediction,'' in \emph{2022 IEEE 38th international conference on data engineering (ICDE)}.\hskip 1em plus 0.5em minus 0.4em\relax IEEE, 2022, pp. 727--740.

\bibitem[Tian et~al.(2022)Tian, Chang, Niu, Song, and Li]{wmlmmi}
Y.~Tian, J.~Chang, Y.~Niu, Y.~Song, and C.~Li, ``When multi-level meets multi-interest: A multi-grained neural model for sequential recommendation,'' in \emph{Proceedings of the 45th international ACM SIGIR conference on research and development in information retrieval}, 2022, pp. 1632--1641.

\bibitem[Wang and Shen(2023)]{imsr}
Z.~Wang and Y.~Shen, ``Incremental learning for multi-interest sequential recommendation,'' in \emph{2023 IEEE 39th International Conference on Data Engineering (ICDE)}.\hskip 1em plus 0.5em minus 0.4em\relax IEEE, 2023, pp. 1071--1083.

\bibitem[Zhang et~al.(2022)Zhang, Yang, Yao, Lu, Feng, Zhao, Chua, and Wu]{re4}
S.~Zhang, L.~Yang, D.~Yao, Y.~Lu, F.~Feng, Z.~Zhao, T.-S. Chua, and F.~Wu, ``Re4: Learning to re-contrast, re-attend, re-construct for multi-interest recommendation,'' in \emph{Proceedings of the ACM Web Conference 2022}, 2022, pp. 2216--2226.

\bibitem[Zheng et~al.(2024)Zheng, Wang, Liu, and Lin]{comorec}
Y.~Zheng, G.~Wang, Y.~Liu, and L.~Lin, ``Diversity matters: User-centric multi-interest learning for conversational movie recommendation,'' in \emph{Proceedings of the 32nd ACM International Conference on Multimedia}, 2024, pp. 9515--9524.

\bibitem[Zhao et~al.(2020)Zhao, Li, Xiao, Deng, and Sun]{catn}
C.~Zhao, C.~Li, R.~Xiao, H.~Deng, and A.~Sun, ``Catn: Cross-domain recommendation for cold-start users via aspect transfer network,'' in \emph{Proceedings of the 43rd International ACM SIGIR Conference on Research and Development in Information Retrieval}, 2020, pp. 229--238.

\bibitem[Fu et~al.(2019)Fu, Peng, Wang, Xu, and Li]{dfrcdr}
W.~Fu, Z.~Peng, S.~Wang, Y.~Xu, and J.~Li, ``Deeply fusing reviews and contents for cold start users in cross-domain recommendation systems,'' in \emph{Proceedings of the AAAI Conference on Artificial Intelligence}, vol.~33, no.~01, 2019, pp. 94--101.

\bibitem[Hsieh et~al.(2017)Hsieh, Yang, Cui, Lin, Belongie, and Estrin]{cml}
C.-K. Hsieh, L.~Yang, Y.~Cui, T.-Y. Lin, S.~Belongie, and D.~Estrin, ``Collaborative metric learning,'' in \emph{Proceedings of the 26th international conference on world wide web}, 2017, pp. 193--201.

\bibitem[Hu et~al.(2013)Hu, Cao, Xu, Cao, Gu, and Zhu]{cdtf}
L.~Hu, J.~Cao, G.~Xu, L.~Cao, Z.~Gu, and C.~Zhu, ``Personalized recommendation via cross-domain triadic factorization,'' in \emph{Proceedings of the 22nd international conference on World Wide Web}, 2013, pp. 595--606.

\bibitem[Mnih and Salakhutdinov(2007)]{tgt}
A.~Mnih and R.~R. Salakhutdinov, ``Probabilistic matrix factorization,'' \emph{Advances in neural information processing systems}, vol.~20, 2007.

\bibitem[Wang et~al.(2021)Wang, Zhuang, Zhang, Wang, Zhou, and He]{lacdr}
T.~Wang, F.~Zhuang, Z.~Zhang, D.~Wang, J.~Zhou, and Q.~He, ``Low-dimensional alignment for cross-domain recommendation,'' in \emph{Proceedings of the 30th ACM international conference on information \& knowledge management}, 2021, pp. 3508--3512.

\bibitem[Li et~al.(2022)Li, Zhao, Zhang, Yu, Cheng, Shu, Kong, and Niu]{recguru}
C.~Li, M.~Zhao, H.~Zhang, C.~Yu, L.~Cheng, G.~Shu, B.~Kong, and D.~Niu, ``Recguru: Adversarial learning of generalized user representations for cross-domain recommendation,'' in \emph{Proceedings of the fifteenth ACM international conference on web search and data mining}, 2022, pp. 571--581.

\bibitem[Bharadhwaj(2019)]{mlcdr}
H.~Bharadhwaj, ``Meta-learning for user cold-start recommendation,'' in \emph{2019 International Joint Conference on Neural Networks (IJCNN)}.\hskip 1em plus 0.5em minus 0.4em\relax IEEE, 2019, pp. 1--8.

\bibitem[Dong et~al.(2020)Dong, Yuan, Yao, Xu, and Zhu]{mamo}
M.~Dong, F.~Yuan, L.~Yao, X.~Xu, and L.~Zhu, ``Mamo: Memory-augmented meta-optimization for cold-start recommendation,'' in \emph{Proceedings of the 26th ACM SIGKDD international conference on knowledge discovery \& data mining}, 2020, pp. 688--697.

\bibitem[Lee et~al.(2019)Lee, Im, Jang, Cho, and Chung]{melu}
H.~Lee, J.~Im, S.~Jang, H.~Cho, and S.~Chung, ``Melu: Meta-learned user preference estimator for cold-start recommendation,'' in \emph{Proceedings of the 25th ACM SIGKDD International Conference on Knowledge Discovery \& Data Mining}, 2019, pp. 1073--1082.

\end{thebibliography}

\vfill

\end{document}